\documentclass[9pt,aps,prd,onecolumn,preprintnumbers,showkeys,nofootinbib]{revtex4}

\usepackage{bm}
\usepackage{natbib}
\usepackage{graphicx}
\usepackage{color}
\usepackage{amsmath,titletoc}

\newcommand{\aj}{{\em Astron.\ J.}}

\renewcommand{\apj}{{\em Astrophys.\ J.}}
\newcommand{\apjl}{{\em Astrophys.\ J.\ Lett.}}
\newcommand{\apjs}{{\em Astrophys.\ J.\ Suppl.}}
\newcommand{\mnras}{{\em Mon.\ Not.\ R.\ Astron.\ Soc.}}

\newcommand{\physrep}{{\em Phys.\ Rep.}}

\newcommand{\jcap}{{\em J.\ Cosmol.\ Astropart.\ Phys.}}
\renewcommand{\prd}{{\em Phys.\ Rev.\ D}}
\renewcommand{\prl}{{\em Phys.\ Rev.\ Lett.}}

\newcommand{\ie}{{\it i.e.\,}}
\newcommand{\eg}{{\it e.g.~}}
\newcommand{\eqn}[1]{Eq.\,(\ref{#1})}
\newcommand{\fig}[1]{Fig.\,\ref{#1}}
\newcommand{\beq}{\begin{equation}}
\newcommand{\eeq}{\end{equation}}
\newcommand{\bdm}{\begin{displaymath}}
\newcommand{\edm}{\end{displaymath}}
\newcommand{\bea}{\begin{eqnarray}}
\newcommand{\eea}{\end{eqnarray}}
\newcommand{\bt}{\begin{tabular}}
\newcommand{\et}{\end{tabular}}

\newcommand{\kv}{{\bf k}}

\newcommand{\qv}{{\bf q}}

\def\d{\delta}
\def\D{\Delta}
\def\Ms{\, h^{-1} \, {\rm M}_{\odot}}

\def\cGpc{\, h^{-3} \, {\rm Gpc}^3}
\def\kMpc{\, h \, {\rm Mpc}^{-1}}

\def\fNL{f_{\rm NL}}

\def\fNLl{f_{\rm NL}^{\rm loc.}}

\def\fNLe{f_{\rm NL}^{\rm eq.}}
\def\dfNLe{\Delta f_{\rm NL}^{\rm eq.}}

\def\dfNLo{\Delta f_{\rm NL}^{\rm or.}}
\def\fNLf{f_{\rm NL}^{\rm fo.}}

\def\O{\mathcal O}

\def\high{\vphantom{\Biggl(}\displaystyle}
\def\w{\omega}
\def\pr{\prime}
\def\nn{\nonumber}
\def\({\left(}
\def\){\right)}
\def\curl{\mathcal}
\def\Blll{B_{l_1l_2l_3}}

\begin{document}


\title{Effects and Detectability of Quasi-Single Field Inflation\\ in the Large-Scale Structure and  Cosmic Microwave Background}

\author{Emiliano Sefusatti}
\email{esefusat@ictp.it}
\affiliation{Institut de Physique Th\'eorique, CEA, IPhT, F-91191 Gif-sur-Yvette, France \&\\
The Abdus Salam International Center for Theoretical Physics, strada costiera, 11, 34151 Trieste, Italy}

\author{James R. Fergusson}\email{jf334@damtp.cam.ac.uk}
\author{Xingang Chen}\email{X.Chen@damtp.cam.ac.uk}
\author{E.P.S. Shellard}\email{E.P.S.Shellard@damtp.cam.ac.uk}
\affiliation{Center for Theoretical Cosmology, Department of Applied Mathematics and Theoretical Physics, University of Cambridge, Wilberforce Road, Cambridge, CB3 0WA, UK}

\begin{abstract}
Quasi-single field inflation predicts a peculiar momentum dependence in the squeezed limit of the primordial bispectrum which smoothly interpolates between the local and equilateral models. This dependence is directly related to the mass of the isocurvatons in the theory which is determined by the supersymmetry. Therefore, in the event of detection of a non-zero primordial bispectrum, additional constraints on the parameter controlling the momentum-dependence in the squeezed limit becomes an important question. We explore the effects of these non-Gaussian initial conditions on large-scale structure and the cosmic microwave background, with particular attention to the galaxy power spectrum at large scales and scale-dependence corrections to galaxy bias. We determine the simultaneous constraints on the two parameters describing the QSF bispectrum that we can expect from upcoming large-scale structure and cosmic microwave background observations. We find that for relatively large values of the non-Gaussian amplitude parameters, but still well within current uncertainties, galaxy power spectrum measurements will be able to distinguish the QSF scenario from the predictions of the local model.   A CMB likelihood analysis, as well as Fisher matrix analysis, shows that there is  also a range of parameter values for which Planck data may be able distinguish between QSF models and the related local and equilateral shapes.   Given the different observational weightings of the CMB and LSS results, degeneracies can be significantly reduced in a joint analysis.
\end{abstract}
\keywords{cosmology: inflation, theory - large-scale structure of the Universe, cosmic microwave background}

\maketitle

\section{Introduction}

Quasi-single field (QSF) inflation models \cite{ChenWang2010A,ChenWang2010B,BaumannGreen2011} are a natural consequence of inflation model-building in string theory and supergravity. In addition to the inflaton field, these models have extra fields with masses of order the Hubble parameter. Such masses are stabilized by the supersymmetry. A distinctive observational signature of these massive fields is a one-parameter family of large non-Gaussianities whose squeezed limits interpolate between the local and the equilateral shape. Therefore, by measuring the precise momentum-dependence of the squeezed configurations in the non-Gaussianities, in principle, we are directly measuring the parameters of the theory naturally determined by the fundamental principle of supersymmetry.

The possibility of detecting a non-Gaussian component in the initial conditions of Early Universe has been the subject of considerable attention in recent years both from an observational perspective and theoretically through inflation model-building \citep{KomatsuEtal2009A, LiguoriEtal2010, Chen2010}. Current constraints from measurements of the bispectrum of temperature fluctuations in the cosmic microwave background (CMB) from the WMAP satellite \citep{KomatsuEtal2011} are still consistent with Gaussianity.  However,  the Planck mission \citep{PLANCK2006} will soon significantly improve the errors on non-Gaussian parameters, leading to strong new constraints or what could be a major breakthrough in cosmology.

The effects of non-Gaussian initial conditions on the large-scale matter and galaxy distributions have been the subject of several studies for more than a decade (see \citep{DesjacquesSeljak2010C} and \citep{LiguoriEtal2010} for recent reviews). The most direct of such effect consists in the additional contribution to the matter bispectrum due to the linearly evolved initial component. Measurements of the galaxy bispectrum in upcoming, large-volume galaxy surveys are expected to improve even  over ideal CMB limits, for any non-Gaussian model \citep{ScoccimarroSefusattiZaldarriaga2004, SefusattiKomatsu2007}.

In addition, the relatively recent discovery of a significant scale-dependent corrections to galaxy bias due to non-Gaussian initial conditions \citep{DalalEtal2008} has led to constraints on the local non-Gaussian parameter $\fNLl$ from current Large-Scale Structure (LSS) data-sets, already comparable to those from the CMB \citep{SlosarEtal2008}. Such bias corrections are present for models where the curvature bispectrum takes large values in the squeezed limit and precisely for this reason, the case of Quasi-Single Field inflation is of particular interest. The results of ref.~\citep{DalalEtal2008} motivated a significant number of further works aimed at a rigorous theoretical description of the effect \citep{SlosarEtal2008, MatarreseVerde2008, McDonald2008, TaruyaKoyamaMatsubara2008, AfshordiTolley2008, DesjacquesSeljakIliev2009, Valageas2009, PillepichPorcianiHahn2010, GiannantonioPorciani2010, NishimichiEtal2010, SchmidtKamionkowsky2010, ShanderaDalalHuterer2011, GongYokoyama2011, CyrRacineSchmidt2011, DesjacquesJeongSchmidt2011A, DesjacquesJeongSchmidt2011B, ScoccimarroEtal2012}. At the same time several groups ran new sets of simulations with non-Gaussian initial conditions of the local \citep{DesjacquesSeljakIliev2009, GrossiEtal2009, PillepichPorcianiHahn2010, NishimichiEtal2010} and other types of models, such as equilateral, orthogonal and folded \citep{WagnerVerdeBoubeker2010, WagnerVerde2012, ScoccimarroEtal2012, GancKomatsu2012, AgulloShandera2012}, as well as generalized local models ({\em e.g.} cubic $g_{NL}$ or scale-dependent $\fNL$ models) \citep{DesjacquesSeljak2010, TseliakhovicHirataSlosar2010, LoVerdeSmith2011, ShanderaDalalHuterer2011}. The picture emerging from such extensive investigations is that the relatively simple expression sufficient to describe the effect of local non-Gaussianity, requires additional corrections to accurately describe models presenting a squeezed limit of the curvature bispectrum different from the local one \citep{DesjacquesJeongSchmidt2011A, ScoccimarroEtal2012}. More generally for such models, the overall correction to bias induced by a generic non-Gaussian model will depend on the halo mass in a non-trivial way and will include a scale-independent component \citep{DesjacquesSeljakIliev2009}. The extension of these predictions to nonlinear bias and to the description of the galaxy bispectrum have been studied in \citep{Sefusatti2009, JeongKomatsu2009B, BaldaufSeljakSenatore2011, GiannantonioPorciani2010} and recently tested in simulations by \citep{NishimichiEtal2010, SefusattiCrocceDesjacques2011} in the context of the local model.

The detectability of the bias correction, both in galaxy and cluster surveys, has also been the subject of several works in recent years \citep{SlosarEtal2008, CarboneVerdeMatarrese2008, Seljak2009, Slosar2009, VerdeMatarrese2009, CarboneMenaVerde2010, CunhaHutererDore2010, SartorisEtal2010, HamausSeljakDesjacques2011, GiannantonioEtal2011, PillepichPorcianiReiprich2011A}. Typical forecasted errors for upcoming galaxy redshift surveys on $\fNLl$, marginalized over cosmological parameters, are of the order of $\Delta\fNLl\sim$ few \citep{CarboneMenaVerde2010, GiannantonioEtal2011}, while cluster surveys including information on the cluster spatial correlations should provide similar results \citep{CunhaHutererDore2010, SartorisEtal2010, PillepichPorcianiReiprich2011A}. Beyond local non-Gaussianity, ref.~\citep{GiannantonioEtal2011} provided forecasts for equilateral and orthogonal models as well, combining information from galaxy and weak lensing observables. When only the 3D galaxy power spectrum is considered, therefore relying mostly on corrections to galaxy bias, the expected errors are of the order of $\dfNLo\simeq 12$ and $\dfNLe\simeq 37$ for orthogonal and equilateral non-Gaussianity, respectively, including CMB priors from the Planck mission. The case of an initial bispectrum described in terms of two parameters, and the possibility of  determining them both, a case complimentary to the one considered here, has been considered by \citep{SefusattiEtal2009} in the context of running non-Gaussianity. In this phenomenological model, theoretically motivated in \citep{Chen2005, ByrnesEtal2010}, the non-Gaussian parameter $\fNL$ presents a scale-dependence parametrized in terms of a running parameter $n_{\rm NG}=\partial\ln \fNL/\partial\ln k$.

Along with these predictions, as already mentioned, other papers derived constraints on non-Gaussian parameters from current observations. In particular, ref.~\citep{SlosarEtal2008} find $-29<\fNLl<70$ at 95\% C.L. combining different data-sets in the Sloan Digital Sky Survey (SDSS), with a dominating contribution from the photometric quasar sample \citep{HoEtal2008}. Ref.~\citep{XiaEtal2010B} finds instead the 2-$\sigma$ limits $3<\fNLl<103$ and  $10<\fNLl<106$ from high-redshift radio sources from the NRAO VLA Sky Survey (NVSS) \citep{CondonEtal1998} and the SDSS quasar sample \citep{RichardsEtal2009}. More recently ref.~\citep{XiaEtal2011} considers the analysis of high-redshift probes for the equilateral and folded models, in addition to the local one, finding, respectively $-480<\fNLe<580$, $-7<\fNLf<373$ and $8<\fNLl<88$ at 95\% C.L. It should be remarked that the model assumed to describe the bias correction in \citep{XiaEtal2011} is quite approximate as it neglects the dependence on the halo mass and further corrections considered for instance in \citep{DesjacquesJeongSchmidt2011A, DesjacquesJeongSchmidt2011B, ScoccimarroEtal2012}, so the limits derived in the equilateral and folded cases are to be considered, as pointed-out by the authors, as limits on ``effective'' non-Gaussian parameters $\fNLe$ and $\fNLf$.

In the first part of this work, we study the effects of QSF models of inflation on Large-Scale Structure with special attention given to the scale-dependent correction to the linear halo bias. The interesting aspect of this correction is its direct dependence on squeezed configurations of the curvature bispectrum. Since, as we will see, the momentum-dependence of these configuration is directly related to one important parameter of the theory, this model constitutes a veritable case study for non-Gaussian effects on halo bias. We will consider the relative importance of both scale-dependent and scale-independent corrections as a function of the halo mass. We will perform a Fisher matrix analysis to assess the detectability of the effect from measurements of the galaxy power spectrum in large-volume redshift surveys and provide a first estimate of the expected uncertainty on the two parameters controlling the initial bispectrum. In addition, we will consider as well the constraints on these parameters expected from measurements of the CMB bispectrum, with particular reference to the upcoming Planck satellite.

This paper is organized as follows. In Section~\ref{sec:qs} we introduce the Quasi-Single Field model of inflation and present the template assumed to describe the predicted curvature bispectrum. In Section~\ref{sec:bsm} we study the effect of the QSF model on the matter bispectrum and skewness, deriving some basic results useful for the following analysis. In Section~\ref{sec:halobias} we consider non-Gaussian corrections to the linear halo bias and show the results for a Fisher matrix analysis of the galaxy power spectrum. In Section~\ref{sec:cmb} we discuss instead the expected constrains from CMB observations and discuss their combination with LSS forecasts. Finally, we present our conclusions in Section~\ref{sec:conclusions}.

\section{Quasi-Single Field Inflation}
\label{sec:qs}

\subsection{Theory}

Inflation model building in supergravity and string theory naturally leads to models of quasi-single field inflation. In this class of multifield models, there is one field direction, the {\em inflaton direction}, that satisfies the slow-roll conditions through either symmetry or fine-tuning, and many other directions, the {\em isocurvaton directions}, that have masses of order the Hubble parameter, $H$. Supersymmetry plays an essential role in determining the masses of these isocurvatons. Without supersymmetry, at the tree level, the coupling of the scalar fields to the space-time curvature would also lead to masses of order $H$, but these masses will run away due to loop corrections, analogous to the situation of the Higgs mass in particle physics. In cosmology, supersymmetry provides the only dynamical mechanism for the masses to maintain this value.
The radiative corrections to the mass from the scalar and fermion loops automatically cancel down to the supersymmetry breaking scale $H$. Therefore, despite  the Hubble parameter being determined by a sector independent of the isocurvatons, and no matter how large or small $H$ is, the mass of light scalars will always trace the value of $H$ through the universal gravitational coupling.
For the inflaton, this mass is the origin of the $\eta$-problem in supergravity inflation models, and needs to be tuned away. For the isocurvatons, they become signatures of supersymmetry in the primordial universe. Finding observational evidence of such scalars constitutes an outstanding theoretical and experimental challenge.

The generic couplings between these isocurvatons and the inflaton have several possible consequences on the primordial density perturbation. The simplest one is the correction to the two-point correlation function, the power spectrum, of the density perturbation. Since the power spectrum is a function of one momentum, the observable effects only appear if such a correction is non-scale-invariant. Furthermore, we expect these effects easily can be degenerate with other types of corrections. The most distinctive signatures of these isocurvatons come from the three or higher-point correlation functions. Unlike the inflaton, the self-interactions of these isocurvaton fields are free of any slow-roll conditions and can be very large. These become the sources of large non-Gaussianities. More importantly, these non-Gaussianities turn out to have very special properties. For example for the scalar three-point function $\langle \Phi^3 \rangle$ in the simplest QSF model, the momentum dependence in the squeezed limit $k_3\ll k_1=k_2$ is given by \cite{ChenWang2010A,ChenWang2010B},
\bea
\langle \Phi^3 \rangle \rightarrow
\left\{
\begin{array}{ccc}
k_3^{-2+\alpha} ~, & & 0< \high{\frac{m^2}{H^2} < \frac{9}{4} + \left( \ln\frac{k_3}{k_1} \right)^{-1} } ~, \\
\high{ k_3^{-3/2} \ln\frac{k_3}{k_1} } ~, & & \high{ \frac{m^2}{H^2} \simeq \frac{9}{4} } ~, \\
\end{array}
\right.
\label{alpha_m}
\eea
where
\bea
\alpha = \frac{1}{2} - \nu ~,
\quad\quad
\nu\equiv \sqrt{\frac{9}{4} - \frac{m^2}{H^2}} ~.
\eea
This is a property of the shape of the non-Gaussianity and is present even if the non-Gaussianity is perfectly scale-invariant. For QSF, the stability of the inflaton requires $m^2\ge 0$, so that $\alpha\ge -1$. As $m^2/H^2 > 9/4$, the isocurvatons gradually become too massive to have significant effects on density perturbations in the absence of sharp features. So we are mainly interested in $-1 \le \alpha \le 1/2$, \ie~$0\le \nu \le 3/2$. Such a momentum dependence lies between that of the equilateral shape ($\alpha=1$) which arises in single field models with large non-Gaussianity or its direct multifield generalization, and that of the local shape ($\alpha=-1$) which arises in multifield models with light isocurvatons $m\ll H$ (see \cite{Chen2010, BartoloEtal2004, CreminelliEtal2011B} for reviews.)

While the detailed dependence of $\alpha$ on the isocurvaton masses may be model-dependent in more general situations, the signature intermediate momentum dependence in the squeezed limit is a robust evidence for the existence of such isocurvatons. This can be seen qualitatively as follows \cite{ChenWang2010A,ChenWang2010B}. The fluctuations of the massive scalars decay after the horizon-exit. For heavy scalars they decay immediately after the horizon-exit, and for lighter scalars they decay more slowly. The scalar interactions, responsible for the large non-Gaussianities, are therefore generated between the horizon scale and the superhorizon scales. The former is responsible for the equilateral-like shapes, and the latter for local-like shapes. As a consistency check, if we look at the special limit of massless scalars, the superhorizon fluctuations do not decay, and we recover the characteristic local shape in the squeezed limit. This momentum dependence can be also seen more quantitatively as follows \cite{BaumannGreen2011},  at least for $\alpha$ close to $-1$ . Ignoring the physics within and near the horizon scale, the squeezed limit of the three-point function can be regarded as the modulation of the two-point function of two short-wavelength modes from a long-wavelength mode. After horizon exit, we know that the amplitude of a massive scalar decays as $\sim a^{-1-\alpha}$ as a function of the scale factor $a$. So the amplitude of the long-wavelength mode has decayed by a factor of $(k_3/k_1)^{1+\alpha}$ by the time the short-wavelength modes start to exit the horizon. The amplitude of the modulation is proportional to the amplitude of the long-wavelength mode. Taking the massless limit as the reference point, at which we know the squeezed limit behavior $\sim k_3^{-3}$ from the simple locality argument \cite{LythRodriguez2005, Starobinsky1985, SasakiStewart1996}, in the massive case we get the momentum dependence $\sim k_3^{-2+\alpha}$.

To illustrate how close these intermediate shapes can get to the local shape but with qualitatively different values of the fundamental parameter, we look at the example of $m = 0.5\,H$. This mass is still of order $H$, qualitatively different from the massless isocurvaton ($m \ll H$) in multifield slow-roll inflation models. But the resulting shape has the momentum dependence $\sim k_3^{-2.9}$, very close to the local shape $k_3^{-3}$ characteristic of the massless isocurvatons. Therefore, how well we can measure the squeezed limit behavior, \ie~the parameter $\nu$, is an important question, if any large non-Gaussianities are discovered.

\subsection{Initial curvature bispectrum}

We will assume throughout the following template for the bispectrum of the Bardeen potential $\Phi$ (with $\Phi=3\,\zeta/5$) \citep{ChenWang2010B}
\beq\label{eq:BispPhi}
B_{\Phi}(k_1,k_2,k_3)=6\, C_\Phi^2\, F(k_1,k_2,k_3)\,,
\eeq
with
\beq\label{eq:tempQsA}
F(k_1,k_2,k_3)=\frac{3^{3/2}}{N_\nu(8/27)}\fNL\frac{N_\nu\left[8 k_1 k_2 k_3/(k_1+k_2+k_3)^3\right]}{(k_1 k_2 k_3)^{3/2}(k_1+k_2+k_3)^{3/2}}\,,
\eeq
where $N_{\nu}$ is the Neumann function of order $\nu$. For simplicity, we will further assume scale invariance for the curvature power spectrum\footnote{Our Fourier transform convention implies the following definitions for power spectra and bispectra $\langle \Phi_{\kv_1}\Phi_{\kv_2}\rangle\equiv \d_D(\kv_{12})\,P_\Phi(k_1)$ and $\langle \Phi_{\kv_1}\Phi_{\kv_2}\Phi_{\kv_3}\rangle\equiv \d_D(\kv_{123})\,B_\Phi(k_1,k_2,k_3)\,$ with $\kv_{ij}\equiv\kv_{i}+\kv_{j}$.},  with the constant $C_\Phi$ defined as $P_{\Phi}(k)\equiv C_\Phi/k^3$. From now on we will generically refer with $\fNL$ to the non-Gaussian amplitude parameter for QSF models as described by the template above. If need be, non-Gaussian parameters for other models will be denoted explicitly with a superscript as, for instance, $\fNLl$ or $\fNLe$ for the local and equilateral shapes, respectively.

As shown by \citep{ChenWang2010B}, the template of \eqn{eq:tempQsA} well reproduces the main features of the bispectrum predicted by QSF models, and, in particular, provides the correct scale-dependence for squeezed configurations as a function of the parameter $\nu$. In fact, in the squeezed limit, \ie $k_3\ll k_1$, $k_2$, the leading order expression for the template becomes, for $\nu\ne 0$,
\beq\label{eq:BphiSqueezedA}
B_{\Phi}(k_1,k_2,k_3)\stackrel{k_3\ll k_1,\,k_2}{\simeq}- \fNL C_\Phi^2 \frac{18\,\sqrt{3}\,\Gamma(\nu)}{4^{\nu}\,\pi \,N_{\nu}(8/27)}\frac{(k_1+k_2)^{3\nu-3/2}}{(k_1\,k_2)^{3/2+\nu}}\frac{1}{k_3^{3/2+\nu}}\,.
\eeq
For the special case $\nu=0$, the limit is given by
\beq\label{eq:BphiSqueezedAnu0}
B_{\Phi}(k_1,k_2,k_3)\stackrel{k_3\ll k_1,\,k_2}{\simeq}- \fNL C_\Phi^2 \frac{36\sqrt{3}}{\pi\,Y_0(8/27)}\frac{\gamma+\ln\left[4\,k_1\,k_2\,/\,(k_1+k_2)^3\right]+\ln\,k_3}{(k_1\,k_2)^{3/2}\,(k_1+k_2)^{3/2}}\frac1{k_3^{3/2}}\,,
\eeq
$\gamma$ being the Euler constant\footnote{For $\nu<0.5$ the next to leading order term goes like $\sim k^{3/2-\nu}$, therefore for $\nu=0$ should be taken into account.}. In particular assuming $k_1=k_2=k_s$ and $k_3=k$, we have  for $\nu\ne 0$,
\beq\label{eq:BphiSqueezedB}
B_{\Phi}(k_s,k_s,k)\stackrel{k\ll k_s}{\simeq}- \fNL C_\Phi^2 \frac{9\,\sqrt{3}\,2^\nu\,\Gamma(\nu)}{\sqrt{2}\,\pi \,N_{\nu}(8/27)}\frac{k_s^{\nu-9/2}}{k^{3/2+\nu}}\,.
\eeq
while for $\nu=0$ we have
\beq\label{eq:BphiSqueezedBnu0}
B_{\Phi}(k_s,k_s,k)\stackrel{k\ll k_s}{\simeq}- \fNL C_\Phi^2 \frac{18\sqrt{3}}{\sqrt{2}\,\pi\,Y_0(8/27)}\(\gamma-\ln\,2+\ln\,k\)\frac{k_s^{-9/2}}{k^{3/2}}\,,
\eeq

\begin{figure*}[t]
{\includegraphics[width=0.9\textwidth]{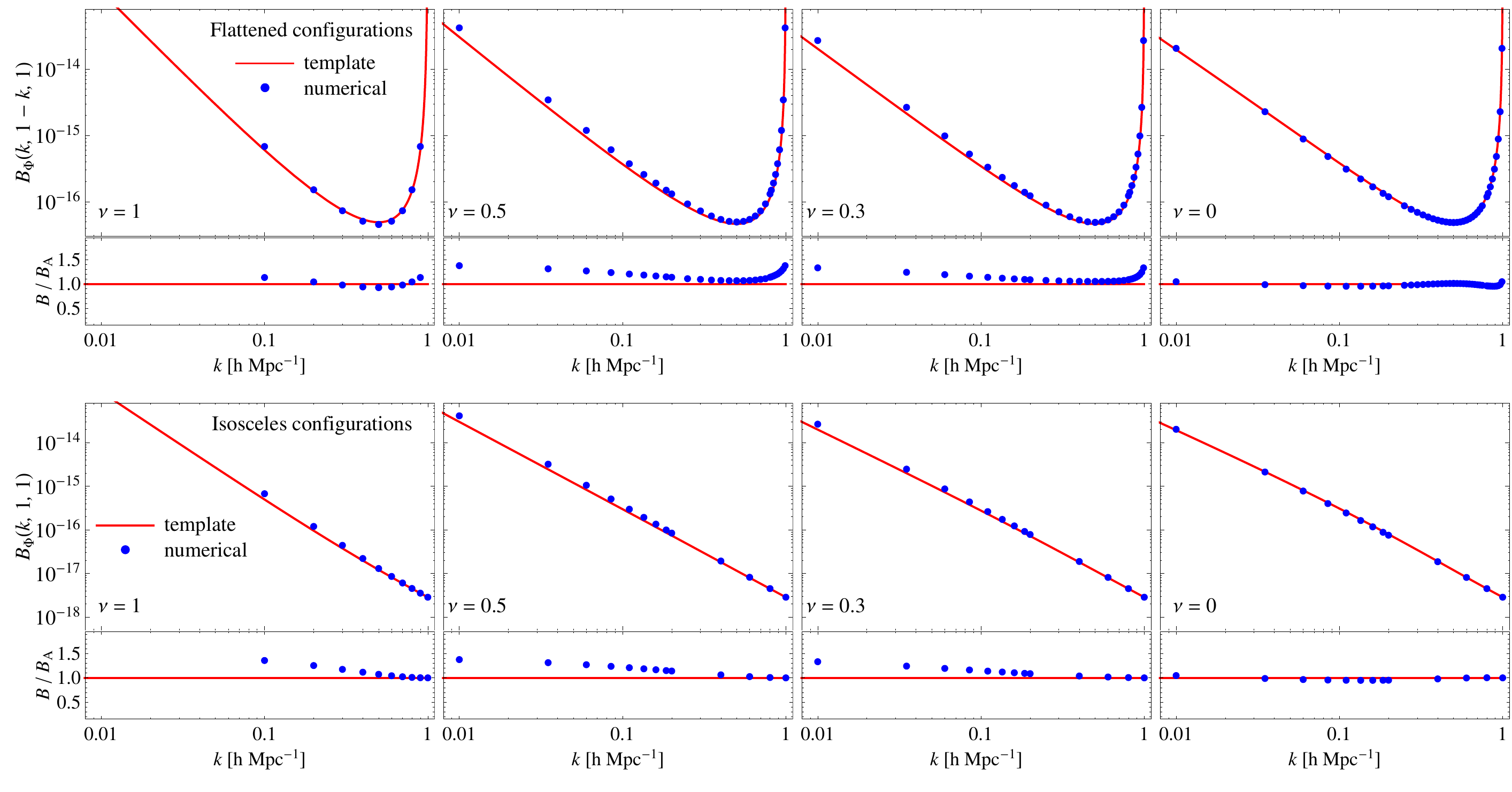}}
\caption{\small {\em Upper panels}: comparison of the bispectrum $B_\Phi$ as described by the template, Eq.~(\ref{eq:tempQsA}), with the numerical evaluation for flattened triangles, $B(k_s,k,k_s-k)$ as a function of $k$ with constant $k_s=1\kMpc$, for $\nu=1$, $0.5$, $0.3$ and $0$ ({\em left to right}). {\em Lower panels}: same comparison for squeezed isosceles triangles, $B(k,k_s,k_s)$ as a function of $k$ with $k_s=1\kMpc$. Notice that a lesser number of points has been evaluated in the $\nu=1$ case.}
\label{fig:templates}
\end{figure*}
In the upper panels of \fig{fig:templates} we compare the squeezed limit of flattened triangular configurations of the curvature bispectrum, that is $B_\Phi(k,k_s-k,k_s)$ with $k_s$ constant as $k\rightarrow 0$, as described by the template above with the numerical evaluation performed in \citep{ChenWang2010B} for $\nu=1$, $0.5$, $0.3$ and $0$.  The lower panels of \fig{fig:templates}  show the same comparison but for squeezed, isosceles triangles, \ie $B_\Phi(k,k_s,k_s)$. It is evident that the template provides an accurate description of the numerical results for $\nu=0$ while it presents an increasing (but asymptotically constant) discrepancy for larger values of $\nu$, of fews tens of percent in the squeezed limit. We remark that this is a minor problem as long as the templates provides the corrected momentum dependence in the squeezed limit. As discussed, this momentum dependence is the most general consequence of the models and is related to the most interesting underlying fundamental physics. In the squeezed limit, this discrepancy simply rescales the definition of $\fNL$. Nonetheless, more accurate comparisons between observations and a specific theoretical model will clearly require a direct evaluation of the bispectrum from the model. So far for the simplest models, direct analytical expressions for the entire shapes are not written in a closed form, and explicit evaluation is only available numerically.

\begin{figure*}[t]
{\includegraphics[width=0.75\textwidth]{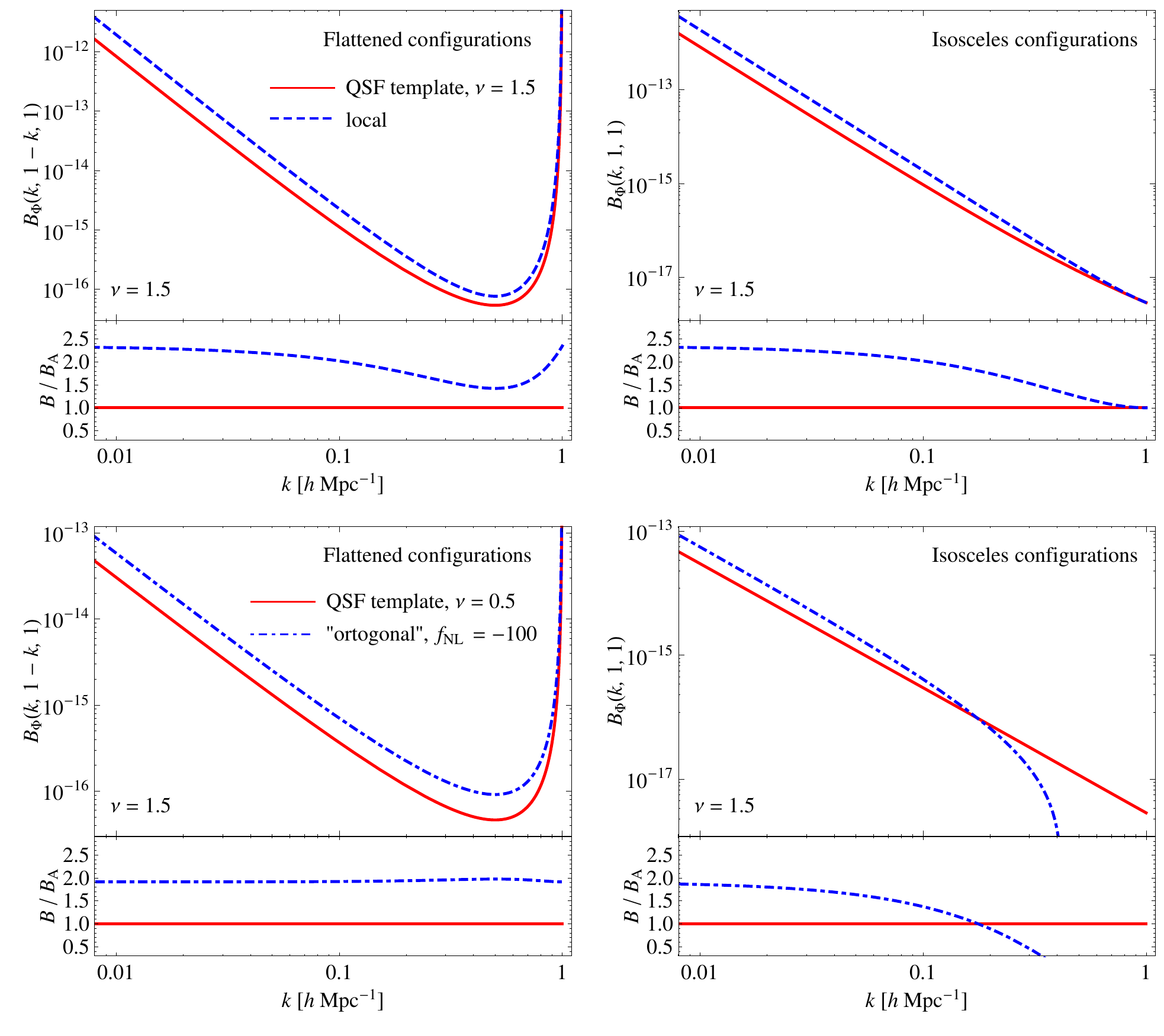}}
\caption{\small {\em Upper panels}: comparison of the QSF template for $\nu=1.5$ with the local shape for flattened squeezed, $B(k_s,k,k_s-k)$ ({\em left panel}) and isosceles, squeezed configurations, $B(k,k_s,k_s)$ ({\em right panel}) assuming $k_s=1\kMpc$. {\em Lower panels}: comparison of the QSF template for $\nu=0.5$ with the orthogonal shape for the same triangular configurations.}
\label{fig:tempLcOr}
\end{figure*}
A ``local limit'' is attained for $\nu=3/2$, leading to $B_\Phi(k_3\ll k_1,k_2)\sim k_3^{-2}$, as for the usual local model. It should be noted, however, that even for $\nu=3/2$ and for squeezed configurations, the amplitude of the QSF template does not coincide with the amplitude of the local model for the same configurations. The upper panels of \fig{fig:tempLcOr} show a comparison between the QSF template for $\nu=3/2$ and the local model bispectrum for both flattened ({\em left}) and isosceles ({\em right}) squeezed configurations. The shape function $F(k_1,k_2,k_3)$ is normalised in such a way to obtain $F(k,k,k)k^6=1$ for any value of $k$, as it happens for the local model, hence the agreement for $k=1\kMpc$ in the plots. For all other configurations this is generically not true. Even for the particular case of $\nu=3/2$ we will denote by $\fNL$ the amplitude parameter for the QSF bispectrum, while we will indicate with $\fNLl$ the analogous parameter for the local model. We find that for squeezed configurations, the two model present the same amplitude if $\fNL\simeq 2.35\fNLl$, with the factor of $2.35$ representing the discrepancy shown in the plots as $k\rightarrow 0$. This relation is useful in comparing the errors on the parameters expected from observations of the galaxy power spectrum since, as we will see in section~\ref{sec:halobias}, the scale-dependent non-Gaussian corrections depend almost exclusively on squeezed configurations. For instance, an expected error of the order of $\Delta \fNLl=5$ would correspond to an error on $\fNL$ of $\Delta \fNL\simeq 12$. More importantly, current LSS constraints $-29<\fNLl<70$ at 95\% C.L. \citep{SlosarEtal2008} translate to $-68<\fNL<164$ for $\nu=1.5$, so that values of $\fNL=150$ are still allowed by galaxy power observations within the 2-$\sigma$ limit. Comparisons with CMB results are less simple, as we will see, since the CMB bispectrum is naturally sensitive to all triangular configurations.

The lower panels of \fig{fig:tempLcOr} compare instead the QSF template for $\nu=1/2$ with the ``orthogonal'' template proposed by \citep{SenatoreSmithZaldarriaga2010} in their Eq.\,(3.2). It should be noted that such template represents a good approximation to the exact orthogonal shape for configurations far from the squeezed limit, and it is therefore a viable choice only for CMB analysis. Its peculiar squeezed limit, in fact, leads to a $1/k$ scale-dependent correction to galaxy bias, not predicted by the true orthogonal shape, but of some phenomenological interest, as it determines an effect on bias intermediate between the local and equilateral models \citep{WagnerVerde2012, ScoccimarroEtal2012}. On the other hand, QSF inflation naturally predicts this kind of behavior. As shown by the figure, for $\nu=0.5$ the asymptotic behavior in the squeezed limit for the two models is the same, with a difference in amplitude of about a factor of two. Notice that for the orthogonal case we assume the same absolute value for $\fNL$ as the QSF model, but opposite sign as the orthogonal bispectrum is negative for such triangles. Far from the squeezed limit, as it is evident in particular from the plot showing isosceles triangles, the two templates are significantly different. For this reason, we will not consider further comparisons between the approximate orthogonal template, but we will instead confront QSF predictions with those of both the local and equilateral shapes.

For all the subsequent calculations in this paper we will assume a flat $\Lambda$CDM cosmological model with the following parameters: $h = 0.719$, $\Delta {\mathcal R}^2 = 2.41\times 10^{-9}$, $\Omega_b = 0.0441$, $\Omega_c = 0.214$ and $n_s=1$, leading to $\sigma_8=0.86$. The matter transfer function is computed with the \texttt{CAMB} code\footnote{\href{http://camb.info}{http://camb.info}}. Notice that, for simplicity we are assuming scale invariance, consistent with the original expression for the bispectrum template proposed by \citep{ChenWang2010B}.

\section{The matter bispectrum and skewness}
\label{sec:bsm}

The most direct effect of primordial non-Gaussianity on large-scale structure is given by its linear contribution to the matter bispectrum. At large-scales, in fact, it is possible to approximately describe the matter bispectrum by its tree-level expression in Eulerian Perturbation Theory (see \citep{BernardeauEtal2002} for a general review on perturbation theory and \citep{LiguoriEtal2010} for the case of non-Gaussian initial conditions). This is given by the sum of the primordial component and the contribution induced by gravitational instability. We have
\beq
B(k_1,k_2,k_3)=B_0(k_1,k_2,k_3)+B_G(k_1,k_2,k_3)\,,
\eeq
where the primordial component, linearly evolved to redshift $z$, is given by
\beq
B_{0}(k_1,k_2,k_3)=M(k_1)M(k_2)M(k_3)B_{\Phi}(k_1,k_2,k_3)\,,
\eeq
with the function
\beq
M(k)\equiv \frac23\frac{D(z)T(k)}{\Omega_{m,0}H_0}k^2\,,
\eeq
expressing the Poisson equation as $\d_\kv=M(k)\,\Phi_\kv$ and $\d_\kv$ representing the matter density contrast in Fourier space. The gravity-induced component is (see \eg \citep{BernardeauEtal2002})
\beq
B_G(k_1,k_2,k_3)=F_2(\kv_1,\kv_2)P_0(k_1)P_0(k_2)+2~{\rm perm.}\,,
\eeq
where $P_0(k)$ is the linear matter power spectrum and the second-order kernel $F_2$ of the perturbative expansion of $\d_\kv$ is given by
\beq
F_2(\kv_1,\kv_2)=\frac57+\frac12\frac{\kv_1\cdot\kv_2}{k_1\,k_2}\left(\frac{k_1}{k_2}+\frac{k_2}{k_1}\right)+\frac27\,\left(\frac{\kv_1\cdot\kv_2}{k_1\,k_2}\right)^2\,.
\eeq

\begin{figure*}[t]
{\includegraphics[width=0.9\textwidth]{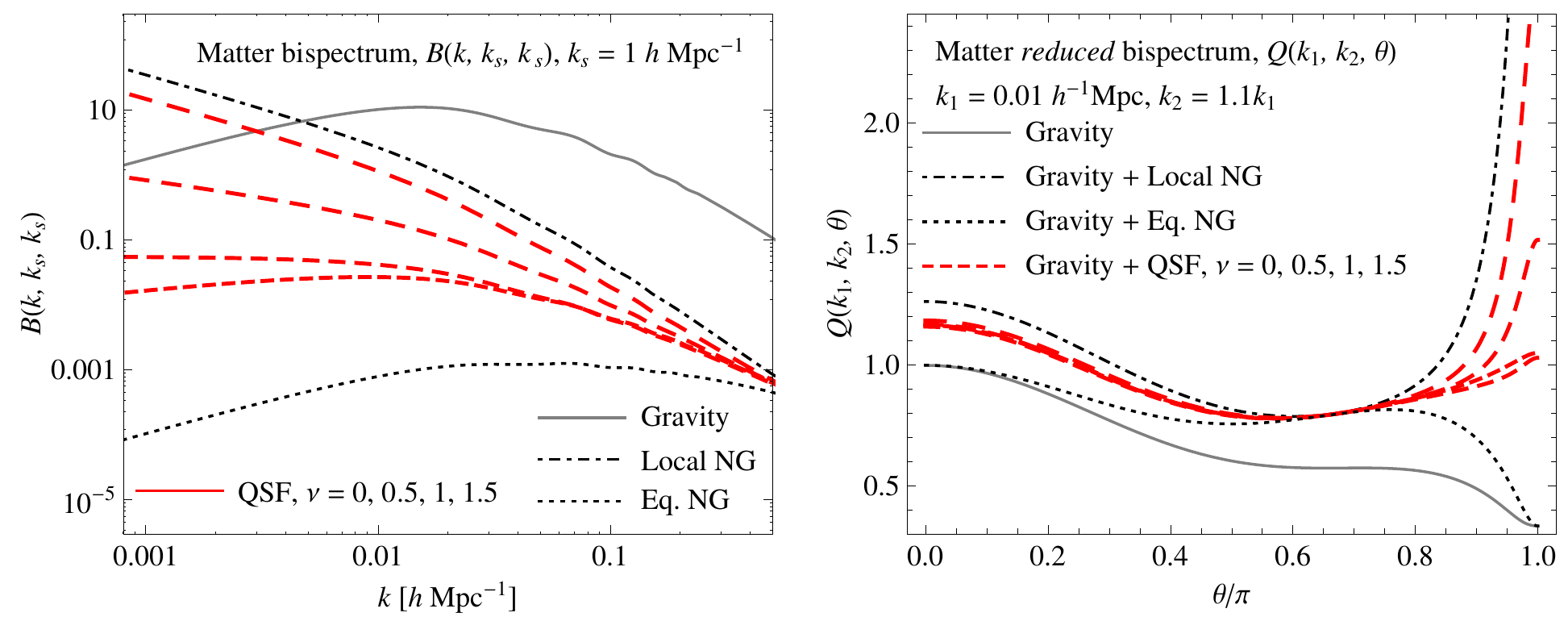}}
\caption{\small {\em Left panel}: comparison of the primordial component to the matter bispectrum due to QSF models ({\em red, dashed curves}) to the component due to the local ({\em black, dot-dashed curve}) and equilateral ({\em black, dotted curve}) non-Gaussian models as well as to the one due to gravitational instability ({\em gray, continuous curve}) for squeezed triangular configurations, \ie  $B(k,k_s,k_s)$ as function of $k$ for fixed $k_s=1\kMpc$. We consider the cases given by $\nu=0$, $0.5$, $1$ and $1.5$ corresponding to the increasingly long-dashed curves from bottom to top.  {\em Right panel}: similar comparison for the {\em reduced} matter bispectrum with fixed $k_1=0.01\kMpc$ and $k_2=1.1\kMpc$ as a function of the angle $\theta$ between $\kv_1$ and $\kv_2$: now the different curves correspond to the same models, but including the gravity contribution.}
\label{fig:bisp}
\end{figure*}
On the left panel of \fig{fig:bisp} we compare the primordial component to the matter bispectrum due to QSF models to the component due to the local and equilateral non-Gaussian models as well as to the one due to gravitational instability for squeezed triangular configurations, \ie $B(k,k_s,k_s)$ as function of $k$ for fixed $k_s=1\kMpc$. In particular we consider for the QSF models the cases given by $\nu=0$, $0.5$, $1$ and $1.5$ corresponding to the increasingly long-dashed curves from bottom to top. We notice how as $\nu$ increases the curves approach the local model, reaching the same dependence on the scale for $\nu=1.5$. The right panel shows the {\em reduced} matter bispectrum, defined as $Q(k_1,k_2,k_3)\equiv B(k_1,k_2,k_3)/[P(k_1)P(k_2)+2~{\rm perm.}]$ where the different curves correspond to the same, different NG models, but where they include in all cases the gravity contribution. For the choice of triangles considered in the limit $\theta\rightarrow 0$, with two sides fixed at close values $k_1=0.01\kMpc$ and $k_2=1.1\kMpc$, this corresponds to the squeezed limit, where different values of $\nu$ lead to distinctly different behaviors.

\begin{figure*}[t]
{\includegraphics[width=0.45\textwidth]{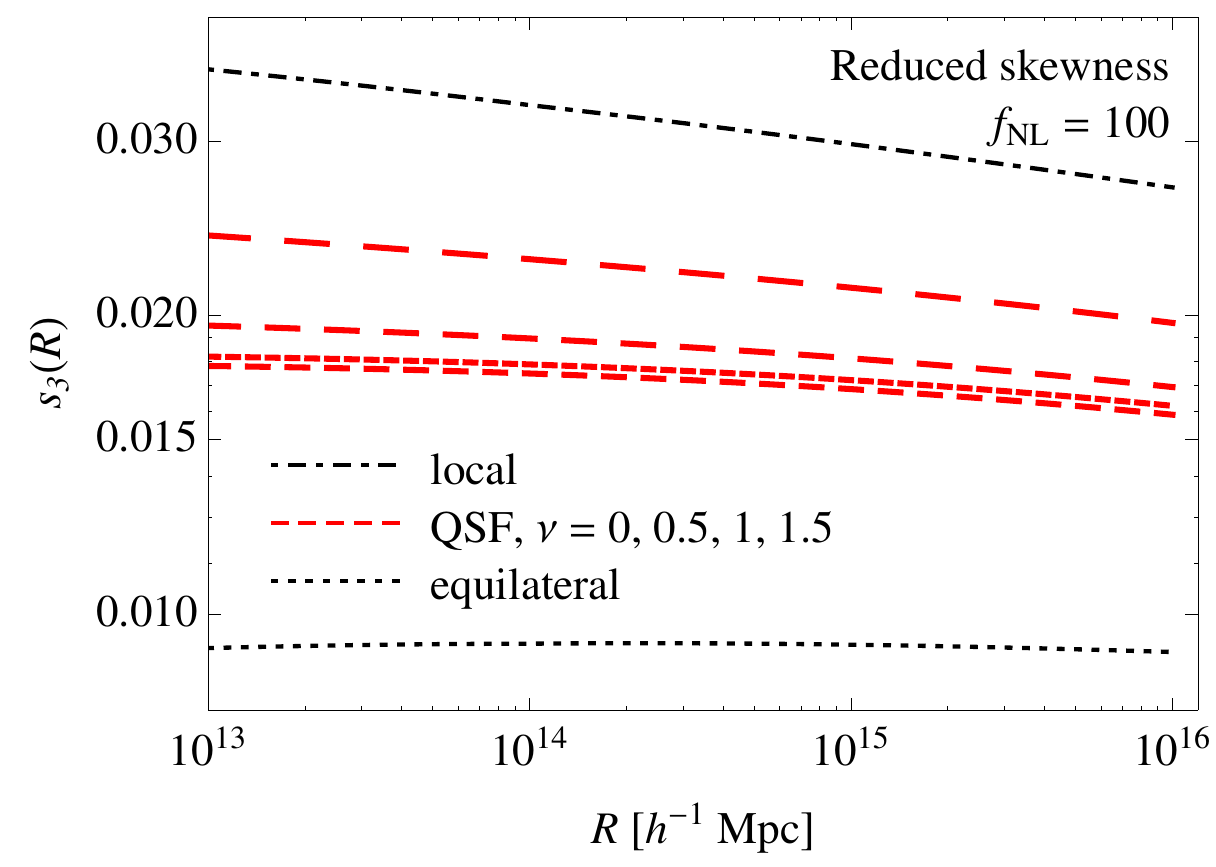}}
\caption{\small Reduced skewness $s_3$, defined in \eqn{eq:skew}, as a function of the halo mass for local ({\em black, dot-dashed curve}), equilateral ({\em black, dotted curve}) and QSF non-Gaussianity with $\nu=0$, $0.5$, $1$ and $1.5$ ({\em red, increasingly long-dashed curves}). Assumes $\fNL=100$, for all models.}
\label{fig:skew}
\end{figure*}
A quantity directly related to the linear matter bispectrum, and useful for the following calculations is given by the reduced skewness of the matter density field $\delta_R$ smoothed on the scale $R$ defined as
\beq\label{eq:skew}
s_3(R)\equiv\frac{\langle\delta_R^3\rangle}{\langle\delta_R^2\rangle^{3/2}}=\frac{\langle\delta_R^3\rangle}{\sigma^3_m}\,,
\eeq
where the third order moment $\langle\delta_R^3\rangle$ is given by the integral
\beq
\langle\delta_R^3\rangle = \int d^3q_1\,d^3q_2\,W_R(q_1)\,W_R(q_2)\,W_R(q_{12})\,B_0(q_1,q_2,q_{12})\,,
\eeq
with $W_R(x)$ being the Fourier transform of a top-hat function. We notice that the reduced skewness $s_3$ does not depend on redshift as well as on the normalization of primordial fluctuations and it is generically mildly depends on the other cosmological parameters.\footnote{In the subsequent calculation we make use of a fit for the reduced skewness of QSF models, function of the parameter $\nu$ and of $x=\delta_c/\sigma_R$ given by $s_3(x,\nu)=\exp\left[c_1(\nu)\ln x+c_2(\nu)(\ln x)^2+c_3(\nu)(\ln x)^3+c_4(\nu)(\ln x)^4\right]$, with $c_1(\nu)=-8.61 - 0.164\,\nu + 0.240\,\nu^2$, $c_2(\nu)=-0.0173 + 0.0419\,\nu - 0.0760\,\nu^2$, $c_3(\nu)=-0.0351 - 0.0176\nu + 0.0298\,\nu^2$ and $c_4(\nu)=0.00511 + 0.00330\,\nu - 0.00513 \nu^2$, accurate at the 1\% level for $0.5<x<14$.} In addition, it is almost constant even with respect to same variable $R$. This evident from \fig{fig:skew} where we plot $s_3(R)$ for local, equilateral and QSF models with $\nu=0$, $0.5$, $1$ and $1.5$ assuming in all cases $\fNL=100$.

The effect of primordial non-Gaussianity on matter correlators is not limited to the large-scale bispectrum, but also involves corrections to the small-scales nonlinear evolution of both power spectrum \citep{TaruyaKoyamaMatsubara2008, WagnerVerdeBoubeker2010, SmithDesjacquesMarian2011} and bispectrum \citep{SefusattiCrocceDesjacques2010, FigueroaEtal2012}. In the case of the bispectrum, such small scale corrections can be significant, but can be directly accessible only via weak lensing measurements.  A proper assessment of the possibility offered by future weak lensing surveys to constrain primordial non-Gaussianity with the measurement of the shear higher-order correlation functions is not yet available. Instead, studies of non-Gaussian effects on galaxy correlators have witnessed recently a great deal of activity, mainly due to the scale-dependent corrections to galaxy bias which we will consider in the next section.

\section{Linear halo bias}
\label{sec:halobias}

Only relatively recently, N-body simulations with local non-Gaussian initial conditions have shown that the bias of dark matter halos receives a significant scale-dependent correction at large scales \citep{DalalEtal2008, DesjacquesSeljakIliev2009, GrossiEtal2009, DesjacquesSeljak2010, PillepichPorcianiHahn2010, WagnerVerde2012, ScoccimarroEtal2012}. Several papers assumed different approaches in the theoretical description of this effect, mostly based on the peak-background split framework or on the theory of peak correlations (see for instance \citep{DesjacquesSeljak2010B} and references therein).

In this work, we will consider the approach of \citep{ScoccimarroEtal2012}, based on the peak-background split argument \citep{BBKS1986, ColeKaiser1989}, tested in N-body simulations assuming different non-Gaussian models for the initial conditions. We assume that the
Eulerian relation, in Fourier space, between the halo density contrast $\delta_h(m)$ for a given halo mass $m$ and the matter density contrast $\delta$ is given, at the linear level, by
\beq
\delta_h(m,\kv)\simeq b_{h}(m,k)\,\delta_\kv\,,
\eeq
with the linear halo bias function
\beq
b_{h}(m,k)=b_{si}(m)+\Delta b_{sd}(m,k)\,,
\eeq
where we distinguish a scale-independent contribution $b_{si}(m)$ from the scale-dependent correction $\Delta b_{sd}(m,k)$.

In this section we will present theoretical results regarding the linear halo bias, keeping in mind that the effects described naturally translate into effects on the linear bias of the galaxy distribution. Galaxy bias can in fact be described as an integral of the halo bias weighted by the halo mass function and by a prescription on how to populate halos with galaxies, the Halo Occupation Distribution (HOD). We will return to this issue in Sec.~\ref{sec:fisher} where we will discuss the detectability of non-Gaussian corrections to the galaxy power spectrum at large-scales.

\subsection{Mass function and scale-independent corrections to bias}

The scale-independent contribution can be derived from the halo mass function $n(m)$  with the usual relation \citep{BBKS1986}
\beq\label{eq:biasPBS}
b_{si}(m)=1+\left.\frac{\partial \ln n(m)}{\partial \delta_l}\right|_{\delta_l=0}\,,
\eeq
under the assumption of the universality of the mass function and Markovianity in the excursion set derivation, with $\delta_l$ representing the large-scale component of the matter fluctuations. In the case of non-Gaussian initial conditions, the mass function receives a correction that leads in turn to a scale-independent non-Gaussian correction. Assuming the non-Gaussian mass function to be described as $n_{NG}(m)=n_G(m)\,R_{NG}(m,\fNL)$ \citep{SefusattiEtal2007} we can write
\beq
b_{si}(m)=b_{G}(m)+\Delta b_{si}(m)\,,
\eeq
where the Gaussian component $b_{G}$ is obtained from \eqn{eq:biasPBS} using the Gaussian mass function $n_G(m)$, while the non-Gaussian correction is given by \citep{DesjacquesSeljakIliev2009}
\beq
\Delta b_{si}(m,\fNL)=\left.\frac{\partial \ln R_{NG}(m,\fNL)}{\partial \delta_l}\right|_{\delta_l=0}\,.
\eeq
We will assume the Sheth \& Tormen \citep{ShethTormen1999} expression for the Gaussian mass function  while for the non-Gaussian relative correction $R_{NG}$ we assume the simple description of \citep{LoVerdeEtal2008} based on the Edgeworth expansion of the non-Gaussian matter probability distribution function in the Press-Schechter framework \citep{PressSchechter1974}. At linear order in $\fNL$ this is given by
\beq\label{eq:Rng}
R_{NG}(m,\fNL)=1 + \frac16\,x\,(x^2 - 3)\, s_3(x) - \frac{1}{6}\, \left(x - \frac1x\right)\,\frac{ds_3(x)}{d\ln x}+{\mathcal O}(\fNL^2)\,,
\eeq
where the variable $x$ is defined as $x(m)\equiv \delta_c/\sigma_m$ with $\delta_c=1.686$ being the linear threshold for spherical collapse and $\sigma_m$ the r.m.s. of the matter perturbations smoothed on the radius $R=[3m/(4\pi \bar\rho)]^{1/3}$, with $\bar\rho$ denoting the mean matter density. In addition, $s_3(x)$ represents the reduced skewness of the initial matter density field, Eq.~(\ref{eq:skew}). To improve the agreement between the measurements of the mass function correction in numerical simulations with the prediction of Eq.~(\ref{eq:Rng}), a scaling parameter $q$ defined by $R_{NG}(x)\rightarrow R_{NG}(q\,x)$ has been considered \citep{GrossiEtal2009, MaggioreRiotto2010C, ParanjapeGordonHotchkiss2011}. We assume here $q=0.91$ as derived in \citep{SefusattiCrocceDesjacques2011} from the simulations of \citep{DesjacquesSeljakIliev2009} (but see also \citep{TinkerEtal2008}, for a similar correction in the context of Gaussian initial conditions).

\begin{figure*}[t]
{\includegraphics[width=0.9\textwidth]{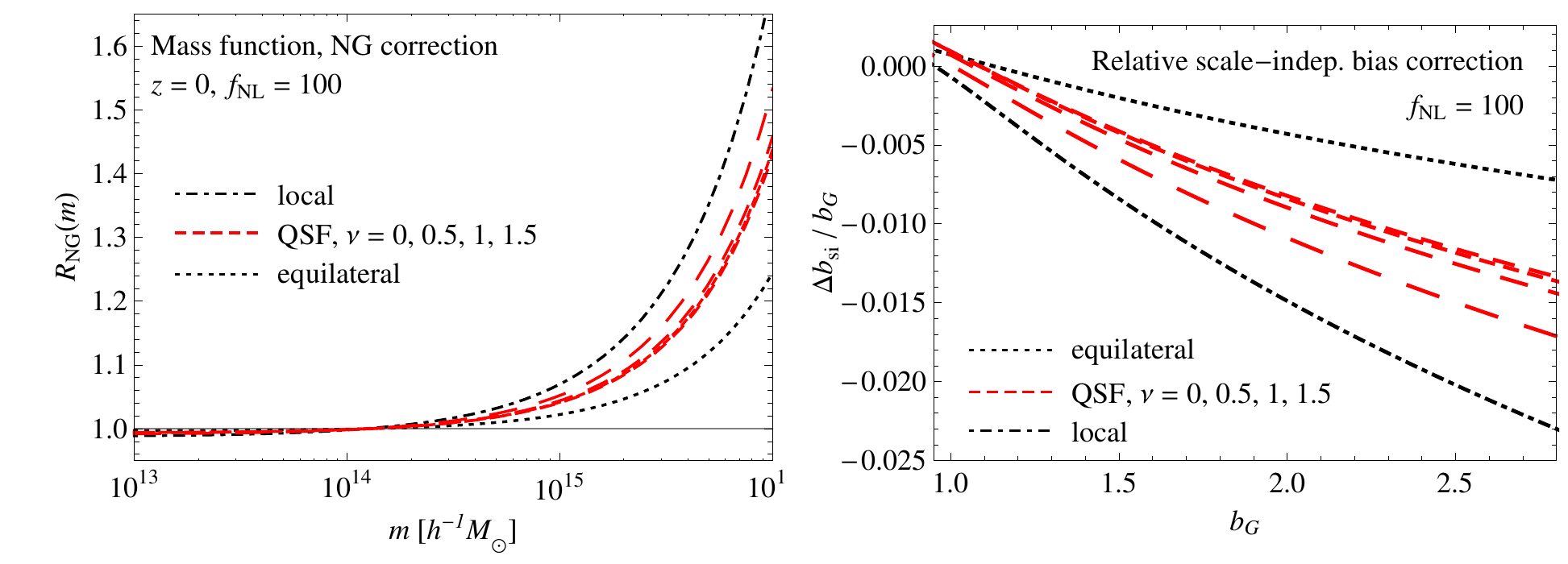}}
\caption{\small {\em Left panel}: non-Gaussian correction to the halo mass function at $z=0$. Different models are denoted as in the left panel. {\em Right panel}: Relative, scale-independent bias correction, $\Delta b_{1,si}/b_{1,G}$, as a function of linear bias $b_{1,G}$.}
\label{fig:mf}
\end{figure*}
In the left panel of \fig{fig:skew} we plot the non-Gaussian correction to the halo mass function at $z=0$. Both the skewness and consequently (in this description) the mass function, at fixed $\fNL$, do not depend strongly on $\nu$, at least for $\nu\lesssim 1$. A mild dependence can be noticed as $\nu$ approaches the limiting value of 1.5. On the right panel of \fig{fig:skew}  we show the {\em relative} scale-independent bias corrections given by $\Delta b_{si}/b_G$ as a function of $b_G$  for the same models and $\fNL=100$. Such corrections are  typically negative and below one percent for moderate values of $b_G$.

\subsection{Scale-dependent corrections to bias}

For the {\em scale-dependent} correction $\Delta b_{1,sd}$ to the linear halo bias due to non-Gaussian initial conditions we assume the following expression from \citep{ScoccimarroEtal2012}
\beq\label{eq:dbsd}
\Delta b_{sd}(m;k,\fNL)= \frac{\delta_c\left[b_{1,G}(m)-1\right]}{2 M(k)}\frac{I_{21}(k,m)}{\sigma^2_m}+\frac{1}{M(k)}\partial_{\ln\sigma_m^2}\left[\frac{I_{21}(k,m)}{\sigma^2_m}\right]\,,
\eeq
where
\beq\label{eq:I21}
I_{21}(k,m)\equiv\frac{1}{P_\Phi(k)}\int d^3q M(q)M(|\qv-\kv|)W_R(q)W_R(|\qv-\kv|)B_\Phi(q,|\qv-\kv|,k)\,,
\eeq
and with $W_R(k)$ a top-hat filter function of radius $R$ corresponding to the halo mass $m$ as $m=(4\pi/3)\bar{\rho}R^3$. This description of the linear bias correction is valid under the assumptions of Markovianity and universality of the mass function which we adopt here for simplicity.

For local non-Gaussianity it is easy to show that in the large-scale limit
\beq
I_{21}(k,m)\stackrel{k\rightarrow 0}{\simeq}4\,\fNL\,\sigma_m^2\,,
\eeq
so that the second term in Eq.~(\ref{eq:dbsd}) vanishes and the first term gives
\beq
\Delta b_{sd}(m;k,\fNL)\simeq 2\,\fNL\,\delta_c\frac{b_{1,G}(m)-1}{M(k)}\,,
\eeq
that is the usual expression first proposed in \citep{DalalEtal2008, SlosarEtal2008} with $M(k)\sim k^2$ for small $k$. The first term of \eqn{eq:dbsd} however generalizes to any non-Gaussian model as in \citep{SchmidtKamionkowsky2010}, while the second term accounts for additional corrections studied in \citep{DesjacquesJeongSchmidt2011A, DesjacquesJeongSchmidt2011B}. In addition, the full result of \citep{ScoccimarroEtal2012} accounts as well for non-Markovian effects and departures from universality of the mass function. Such effects will have to be taken into account in a proper comparison with numerical simulations results but we can neglect them here as they will not affect our results.

In the case of the QSF model as described by the template for the curvature bispectrum in \eqn{eq:tempQsA}, from the squeezed limit, \eqn{eq:BphiSqueezedA}, we can derive the large-scale approximation of the integral $I_{21}(k,m)$, which is given, for $\nu\ne 0$, by
\beq\label{eq:I21app}
I_{21}(k,m)\stackrel{k\rightarrow 0}{\simeq}-\,\fNL\,\frac{18\,\sqrt{3}\,\Gamma(\nu)}{2^{\nu-3/2}\,\pi \,Y_{\nu}(8/27)}\,k^{3/2-\nu}\,\Sigma_{3/2-\nu}\,,
\eeq
where, following the notation of \citep{ScoccimarroEtal2012},
\beq
\Sigma_n(R)\equiv\int d^3 q\, P(q)\, W_R^2(q)\,q^{-n}\,.
\eeq
In the case $\nu=0$, we have
\beq\label{eq:I21appnu0}
I_{21}(k,m)\stackrel{k\rightarrow 0}{\simeq}\,\fNL\,\frac{36\,\sqrt{3}}{2^{3/2}\,\pi \,Y_{0}(8/27)}\,k^{3/2}\,\left[\left(\gamma+\ln \frac{k}2\right)\,\Sigma_{3/2-\nu}-\int d^3 q\,\frac{\ln q}{q^{3/2}} P(q)\, W_R^2(q)\,\right]\,.
\eeq

Since at large scales $M(k)\sim k^2$, the expected scale-dependent correction to halo bias presents in this case the behavior
\beq
\Delta b_{sd}(k)\sim k^{-1/2-\nu}\,.
\eeq
Over the allowed range of values of $\nu$, $0\le \nu\le 3/2$, such corrections will therefore interpolate between $\Delta b_{sd}(k)\sim k^{-1/2}$ and $\Delta b_{sd}(k)\sim k^{-2}$, the latter corresponding to the effect of local NG.

\begin{figure*}[t]
{\includegraphics[width=0.9\textwidth]{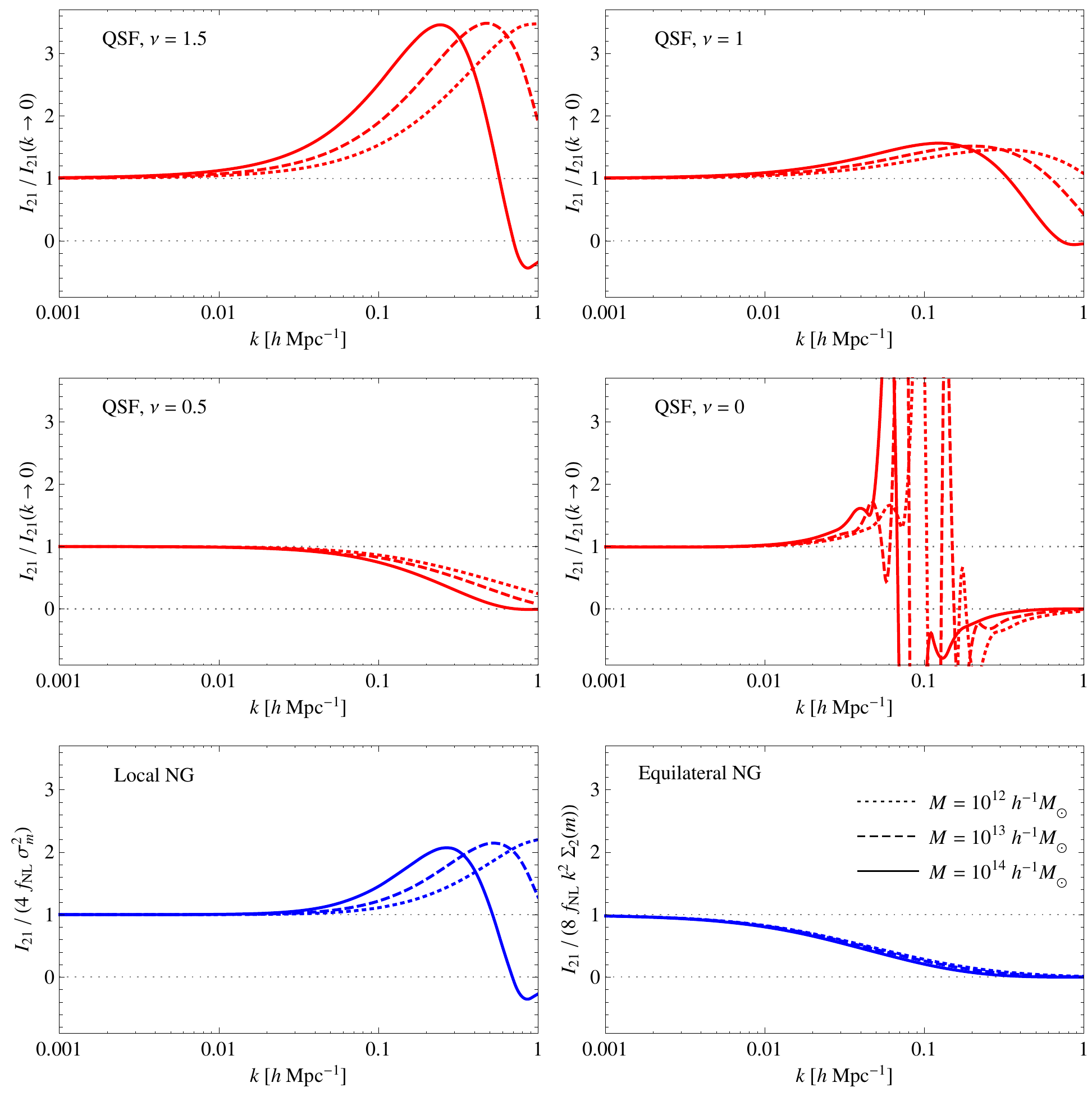}}
\caption{\small Comparison of the full calculation of the function $I_{21}$, Eq.~(\ref{eq:I21}), with its asymptotic value as $k\rightarrow 0$ for the QSF model with $\nu=1.5$ ({\em top left panel}), $\nu=1$ ({\em top right panel}), $\nu=0.5$ ({\em central left panel}) and $\nu=0$ ({\em central right panel}). The lower panels show the same quantities for the local ({\em left}) and equilateral ({\em right}) models. Dotted, dashed and continuous curves correspond respectively to halo masses $m=10^{12}$, $10^{13}$ and $10^{14}\Ms$. Notice that the plotted ratio is independent of redshfit. The noisy results for the $\nu=0$ case are due to the vanishing of the asymptotic expression, \eqn{eq:I21appnu0} due to the canceling of the two terms on the r.h.s.}
\label{fig:I21}
\end{figure*}
In \fig{fig:I21} we show the quantity $I_{21}(m,k)$, \eqn{eq:I21}, as a function of $k$ evaluated for different halo masses and divided by its asymptotic value, Eq.s\,(\ref{eq:I21app}) and (\ref{eq:I21appnu0}) (see also Fig.\,2 and 3 in \citep{ScoccimarroEtal2012}). We notice that different masses correspond to specific dependence on scale for intermediate values of $k$. In the case of the QSF model for $\nu=1.5$ this aspect is more evident than in the otherwise similar local model. It is evident that the asymptotic approximation is able to describe the effect only at the very largest scales, while the full evaluation of the integral $I_{21}(m,k)$ is required already for wavenumbers above $k\sim 0.01\kMpc$.

\begin{figure*}[t]
{\includegraphics[width=0.9\textwidth]{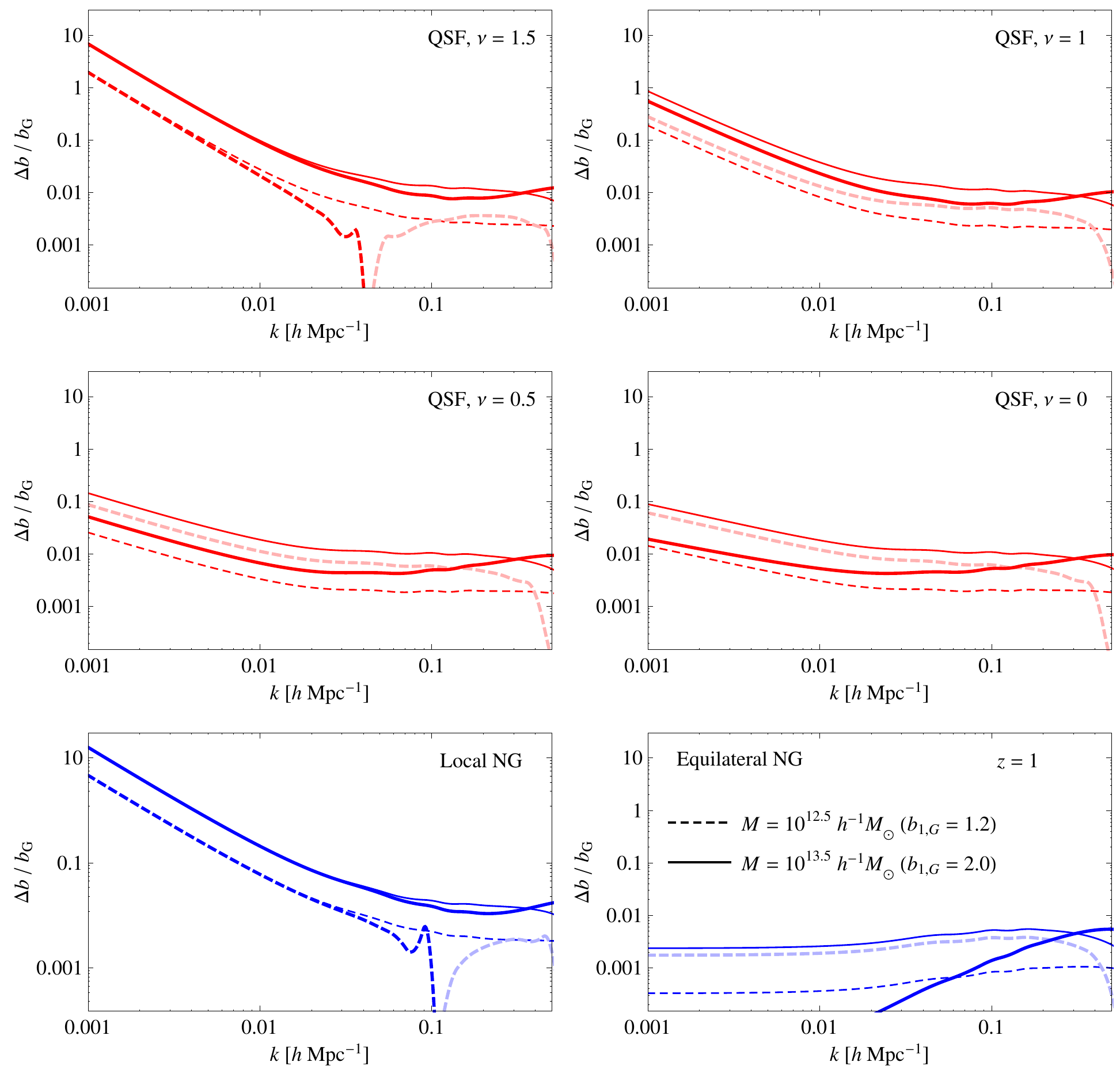}}
\caption{\small Relative scale-dependent correction $\Delta b_{sd}(k)/b_G$ ({\em thick dashed and thick continuous curves}) as a function of $k$ for two representative halo masses $m=10^{12.5}\Ms$ ({\em dashed ones}) and $m=10^{13.5}\Ms$ ({\em continuous ones}), among which the light-colored ones denote negative values. The choice of the models shown is the same as \fig{fig:I21}. Thin dashed and thin continuous curves correspond to sole contribution due to the first term on the r.h.s. of Eq.~(\ref{eq:dbsd}).}
\label{fig:db}
\end{figure*}
In \fig{fig:db} we finally show, with thick curves, the relative scale-dependent correction $\Delta b_{sd}(k)/b_G$ with $b_{sd}$ given by \eqn{eq:dbsd} as a function of $k$ for two representative halo masses $m=10^{12.5}$  and $10^{13.5}\Ms$. Thin curves correspond instead to sole contribution due to the first term on the r.h.s. of \eqn{eq:dbsd} while light-colored curves denotes negative values. It is evident, in the first place that the case $\nu=1.5$ is indeed very close to the local NG case, up to the mentioned numerical factor, both in term of scale-dependence and as a function of mass. As we consider lower values of $\nu$ the dependence on mass, and, in particular, the contribution on the second term on the r.h.s. of \eqn{eq:dbsd} becomes relevant, particularly for the low mass examples. In fact, already for $\nu=1$, while in the high mass case such additional corrections amounts to a overall decrease in the amplitude of the scale-dependent correction, for the low mass case, they induce a change in the {\em sign} of the correction itself, albeit keeping a similar absolute amplitude. We should remark that this sort of behavior has never been tested in simulations for this specific model, and a dedicated study is clearly required. However, as already mentioned, comparisons with different non-local models (\eg \citep{DesjacquesJeongSchmidt2011A, ScoccimarroEtal2012}) indicate that \eqn{eq:dbsd} does describe correctly the mass dependence of the halo bias corrections to the extent allowed by the error on current numerical results. If this description will find further confirmation, possibly for the specific case of the QSF model, the results shown of \fig{fig:db} would be particularly interesting in terms of observational constraints. One can imagine, in fact, the possibility of combining the measurements of the power spectrum for different galaxy populations where we expect a correction different in sign, allowing in principle for a significant reduction of the degeneracy between the non-Gaussian parameters and the linear Gaussian bias (see \eg \cite{Seljak2009,HamausSeljakDesjacques2011} for recent works in this direction). The other panels of \fig{fig:db} referring to the QSF model, show that a similar situation is recovered for even smaller values of $\nu$, as for instance $\nu=0.5$ corresponding to a scale-dependent correction $\Delta b_{sd}\sim 1/k$.

\subsection{Fisher matrix}
\label{sec:fisher}

We perform a Fisher matrix analysis in order to determine the expected, simultaneous constraints on both the $\fNL$ and $\nu$ parameters from future galaxy surveys. Similar forecasts, in terms of the single $\fNL$ parameter, have been performed, for the local model alone in \citep{DalalEtal2008, CarboneVerdeMatarrese2008, CarboneMenaVerde2010, CunhaHutererDore2010, SartorisEtal2010} for upcoming galaxy and clusters surveys. The cases of equilateral and orthogonal non-Gaussianity, in addition to the local model, have been studied by \citep{GiannantonioEtal2011} and \citep{PillepichPorcianiReiprich2011A}, focusing respectively on the  EUCLID survey \citep{LaureijsEtal2011} (combining weak lensing with photometric and spectroscopic data) and the eRosita, X-ray cluster survey \citep{PredehlEtal2010}. In these works, the effects of non-local models include scale-dependent corrections to halo bias, described by expressions analogous to \eqn{eq:dbsd}. Running non-Gaussianities described by an additional running parameter $n_{NG}$, such as those for both the local and equilateral shapes \citep{Chen2005, ByrnesEtal2010}, also depend on two parameters. Forecasts for these models have been considered in \citep{SefusattiEtal2009, LoVerdeEtal2008, GiannantonioEtal2011}. Some examples of these running non-Gaussianities may also alter the momentum dependence in the squeezed limit of the bispectrum. But the difference between this case and the QSF inflation is clear. The former is caused by the running of non-Gaussianity and the latter by the shape of non-Gaussianity. For QSF inflation, even if $3/2-\nu$ is not much less than one and so the deviation from the local shape is significant, the bispectrum can be still scale-invariant. But for running Gaussianities, such a momentum dependence corresponds to a case with very strong overall scale-dependence.

Our goal is to provide an estimate of the possibility to constrain the parameter $\nu$ assuming a positive detection of non-Gaussianity by Planck. For this purpose we consider two large-volume surveys comparable in size and redshfit range to EUCLID and LSST. Such surveys are already expected to provide constraints to $\fNLl$ \citep{CarboneMenaVerde2010, GiannantonioEtal2011} from measurements of the galaxy power spectrum comparable to those expected by Planck \citep{YadavKomatsuWandelt2007, YadavEtal2008, FergussonShellard2007}. Being this a first, indicative assessment, we will assume a simplified description, characterized simply by the field of view, the redshift range (\ie the volume) and the expected linear (Gaussian) bias $b(z)$ as a function of redshift. Since we expect most of the signal to come from the largest scale probed, we do not expect a significant impact of shot-noise (see, for instance, \citep{CarboneMenaVerde2010, GiannantonioEtal2011}). On the other hand, shot-noise should not affect, by design, the BAO analysis, the primary target of these missions. We will nevertheless include a shot-noise contribution to the power spectrum variance as detailed below. The photometric or spectroscopic nature of redshfit observations is assumed to play a negligible role as, again, the determination of large-scale power does not require high precision in the radial galaxy positions.

For our analysis we assume a (``EUCLID-like'') survey, denoted as V1, with a 20,000 deg$^2$ field of view and a redshift range of $0.4<z<2$, and a galaxy population with fiducial (Gaussian) linear bias parameter given by $b_G(z)=\sqrt{1+z}$ \citep{OrsiEtal2010}. This would allow a comparison with the results of \citep{GiannantonioEtal2011} for the local and equilateral models. As opposed to \citep{GiannantonioEtal2011}, which considers 12 equally populated redshift distributions, we will simply compute the Fisher matrix information in redshift bins of size $\Delta z=0.1$. As an exercise we extend this galaxy population, described by the same bias evolution $b_G(z)$, to a larger (``LSST-like") volume, denoted as V2, given by a field of view of 30,000 deg$^2$ and redshift range $0.3<z<3.8$. This corresponds to the second example of \citep{CarboneMenaVerde2010}, although with a lower (therefore more conservative) fiducial value for the linear bias.

We consider a two-dimensional Fisher matrix ${\mathcal F}_{a,b}$ for the parameters $p_a=\{\fNL,\nu\}$ obtained as a sum over all $N_z$ redshift bins of the three-dimensional matrix $\widetilde{{\mathcal F}}_{\alpha,\beta}$ for the parameters $p_\alpha=\{\fNL,\nu,b_G\}$, marginalized over the linear bias parameter $b_G$. Since the scale-dependent correction $\Delta b_{sd}$ depends linearly on the product $\fNL\,(b_G-1)$, we can expect a significant degeneracy with $\fNL$ and a marginalization over the value of $b_G$ should be taken into account. Such marginalization is performed in \citep{GiannantonioEtal2011} but not in \citep{CarboneMenaVerde2010}, where, on the other hand, only the very largest scales are considered for the analysis of the local model alone.  The matrices are defined respectively as
\beq
{\mathcal F}_{ab}\equiv\sum_{i=1}^{N_z}\left[\left(\widetilde{{\mathcal F}}^{-1}(\bar z_i)\right)_{ab}\right]^{-1}\,,
\eeq
and
\beq\label{eq:fisherLSS}
\widetilde{{\mathcal F}}_{\alpha\beta}\equiv\sum_{k_j=k_{min}(\bar z_j)}^{k_{\rm max}(\bar z_j)}\frac{\partial\,P_g(k_j,\bar z_i)}{\partial\, p_a}\frac{\partial\,P_g(k_j,\bar z_i)}{\partial\, p_b}\frac{1}{\Delta P_g^2(k_j,\bar z_i)}\,,
\eeq
where the sum runs over the available wavenumbers from the fundamental frequency of the redshfit bin $k_{min}\equiv 2\pi/V^{1/3}(z)$, $V(z)$ being the bin volume to $k_{\rm max}$, in steps of $k_{min}$.

The expression for the galaxy power spectrum is given by
\bea
P_g(k,z) & = & b^2(z)\,P(k,z)\,\nonumber\\
 & = & [b_{G}^2(z)+2\,b_{G}(z)\,(\Delta b_{si}(z)+\Delta b_{sd}(k,z))]\,P(k,z)+{\mathcal O}(\fNL^2)\,
\eea
where we include only linear corrections in $\fNL$. It should be noted that primordial non-Gaussianity modifies both the amplitude and the form of the galaxy power spectrum and a large range in $k$ helps reducing degeneracies between non-Gaussian parameters and bias. Following \citep{GiannantonioEtal2011}, we choose $k_{\rm max}=\pi/(2\,R)$ with the scale $R$ obtained from the redshift-dependent equality $\sigma_R^2(z)=0.48$ constant. At $z=0$ this choice provides, for our cosmology, $k_{\rm max}=0.15\kMpc$, as in \citep{GiannantonioEtal2011} and it implies the range of values $0.2\lesssim k_{\rm max} \lesssim 0.5$ for $0.5<z<2$. We will later discuss how our results depend on this choice, comparing them with those obtained from a more conservative assumption corresponding to $k_{\rm max}=0.075\kMpc$ at $z=0$. For the matter power spectrum $P(k,z)$ we assume the linear expression and consequently ignore nonlinear corrections at small scales. While on one side we do expect such corrections to be there for $k$ close to $k_{\rm max}$, the final results do not strongly depend on this assumption since we neglect them as well in the power spectrum variance $\Delta P_g^2$. In fact, in our approximation, their inclusion would only reduce the impact of shot-noise on $\Delta P_g^2$. The range in wavenumbers is, on the other hand, well within reach of accurate predictions in nonlinear perturbation theory (see {\em e.g.} \citep{CrocceScoccimarro2008}). Notice that, for simplicity, we also ignore possible corrections due to nonlinear bias. This is in part justified by the small value of the quadratic bias parameter corresponding to the linear one $b_G(z)$ (typically $|b_2/b|\lesssim0.35$) that can be derived by the expressions of \cite{ScoccimarroEtal2001A}.

The power spectrum variance is given by (see \eg \citep{SefusattiEtal2009})
\beq
\Delta P_g^2(k,z)\equiv\frac{k_f^2}{2\pi\,k^2}P_{tot}^2(k,z)\left[1+\frac{4\,\Delta b_{NG}(k,z)\,P(k,z)}{P_{tot}(k,z)}\right]+{\mathcal O}(\fNL^2)\,,
\eeq
which accounts for non-Gaussian corrections to the bias and where $P_{tot}$ is the Gaussian galaxy power spectrum including shot-noise,
\beq
P_{tot}(k,z)=P_g(k,z)+\frac1{(2\pi)^3\,\bar n(\bar z_i)}\,.
\eeq

Since we are describing the galaxy population simply in terms of the value of the linear bias we derive the mean number density $\bar n(\bar z_i)$ in the redshift bin $\bar z_i$ making the quite drastic but simplifying assumption that each dark matter halo of mass above a certain minimal mass $M_{min}(\bar z_i)$ contains a single galaxy. The value of $M_{min}(\bar z_i)$ is determined imposing the relation
\beq
b_G(\bar z_i)\equiv\sqrt{1+\bar z_i}=\int_{M_{min}(\bar z_i)} n_G(m,\bar z_i)\,b_{G,h}(m,\bar z_i)\,m\,dm\,,
\eeq
where we assume the Sheth-Tormen \citep{ShethTormen1999} expressions for the Gaussian halo mass function $n_G(m,z)$ and linear halo bias $b_{G,h}(m,z)$. The galaxy density is then obtained as
\beq
\bar n(\bar z_i)=\int_{M_{min}(\bar z_i)} n(m,\bar z_i)\,dm\,.
\eeq
We notice that since the halo bias $b_{G,h}(m,\bar z_i)$ weights more massive halos with respect to smaller mass halos, this procedure tends to overestimate the value of $M_{min}$ and therefore underestimates the value of the density when compared to the same calculation performed assuming the proper Halo Occupation Distribution (HOD) for the galaxy sample. In addition, the definition of the minimal mass $M_{min}$ allows us, perhaps improperly, to define a mean or ``characteristic mass'' for the halo population of the bin at $\bar z_i$ given by
\beq
\bar m_i=\int_{M_{min}(\bar z_i)} n(m,\bar z_i)\,m\,dm\,.
\eeq
We assume the mean mass $\bar m(\bar z_i)$ to evaluate the integral $I_{21}(k,m,z)$, \eqn{eq:I21}, and its derivatives for each bin. Notice that in the marginalization over the bias, {\em
i.e.} in the derivative $\partial P_g/\partial b_G$, we not only consider the explicit dependence on the parameter $b_G$ but derive as well {\em all} mass-dependent quantities like $\sigma_m$ or the integrals as $I_{21}$ as $\partial/\partial b_G=[\partial b_{G,h}(m,z)/\partial m]^{-1} (\partial/\partial m$). Obviously, the correct procedure would have involved integrating all bias corrections over the proper range of halo masses and the proper HOD. However, such a drastic solution is still more conservative than neglecting altogether the marginalization over the mass-dependence of such corrections as done in previous similar analysis in the literature. Ref.~\citep{GiannantonioEtal2011} correctly points out that while the mass-dependence of such corrections is relevant in the case of the equilateral model, predictions could not be properly tested in simulations yet for this specific model (see \eg \citep{ScoccimarroEtal2012}) and they restrict their expressions for $\Delta b_{sd}$ to its asymptotic value at small $k$. However, we assume the full expression in \eqn{eq:dbsd} to be valid, noticing that such issues are less important for our model when $\nu\ge 0.5$ and that after marginalization over bias (and mass), our results are, as we will see, consistent with those of \citep{GiannantonioEtal2011} for both the local and equilateral models.

Finally, we do not consider a full marginalization over cosmological parameters. Forecasted constraints on the local $\fNLl$ parameter, from measurements of the galaxy power spectrum marginalized over the cosmology (with priors from the Planck CMB power spectrum) have been studied in \citep{CarboneMenaVerde2010} finding an increase in the error $\Delta\fNLl$ of about 30\% for a EUCLID-like experiment. Ref. \citep{GiannantonioEtal2011} compares instead errors on different NG models marginalized on cosmology with and without Planck priors. They find, as one can expect,  a particularly strong degeneracy of the equilateral $\fNLe$ parameter with cosmological ones, due to the lack of a  scale-dependence in the halo bias correction. We might therefore expect an larger impact of the uncertainty on cosmological parameters when low value of $\nu$ are considered.

\begin{table}[t]
\centering
\begin{tabular}[t]{l||c|c||c|c||c|c||c|c||c|c||c|c||c|c||c|c}
\hline
\hline
  \multicolumn{17}{l}{V1 survey, $V=103\cGpc$, $0.5<z<2$}\\
  \hline
 &  \multicolumn{8}{c||}{$k_{\rm max}=0.15\kMpc$ at $z=0$} & \multicolumn{8}{c}{$k_{\rm max}=0.075\kMpc$ at $z=0$}\\
  \hline
 & \multicolumn{2}{c||}{$\fNL=0$} & \multicolumn{2}{c||}{$\fNL=50$} & \multicolumn{2}{c||}{$\fNL=100$} & \multicolumn{2}{c||}{$\fNL=150$}  & \multicolumn{2}{c||}{$\fNL=0$} & \multicolumn{2}{c||}{$\fNL=50$} & \multicolumn{2}{c||}{$\fNL=100$} & \multicolumn{2}{c}{$\fNL=150$}    \\
\hline
 &$\Delta\fNL$ & $\Delta\nu$ &$\Delta\fNL$ & $\Delta\nu$ & $\Delta\fNL$ &$\Delta\nu$ & $\Delta\fNL$ &$\Delta\nu$  &$\Delta\fNL$ & $\Delta\nu$ &$\Delta\fNL$ & $\Delta\nu$ & $\Delta\fNL$ &$\Delta\nu$ & $\Delta\fNL$ &$\Delta\nu$   \\
\hline
Local & $2.9$& -& $3.9$& -& $3.9$&-&  $4.1$ & -  & $3.1$& -& $4.1$& -& $4.3$&-&  $4.6$ & -\\
\hline
$\nu=1.5$ & $7.3$ & - & $36$&$0.17$&$37$&$0.09$&$14$&$0.03$ & $7.6$ & - & $49$&$0.23$&$53$&$0.13$&$22$&$0.05$  \\
\hline
$\nu=1.0$ & $18$ & - & $32$&$0.22$&$32$&$0.11$&$32$&$0.07$ & $95$ & - & $130$&$0.35$&$130$&$0.18$&$130$&$0.12$  \\
\hline
$\nu=0.5$ & $13$ & - & $42$&$1.86$&$41$&$0.92$&$41$&$0.61$  & $69$  & - & $140$&$3.0$&$140$&$1.5$&$140$&$0.98$  \\
\hline
Equilateral  & $18$& -& $18$& -& $18$&-&  $18$ & - & $141$& -& $139$& -& $137$&-&  $135$ & - \\
\hline
\hline
  \multicolumn{17}{l}{V2 survey, $V=390\cGpc$, $0.4<z<3.6$}\\
  \hline
 &  \multicolumn{8}{c||}{$k_{\rm max}=0.15\kMpc$ at $z=0$} & \multicolumn{8}{c}{$k_{\rm max}=0.075\kMpc$ at $z=0$}\\
  \hline
 & \multicolumn{2}{c||}{$\fNL=0$} & \multicolumn{2}{c||}{$\fNL=50$} & \multicolumn{2}{c||}{$\fNL=100$} & \multicolumn{2}{c||}{$\fNL=150$}  & \multicolumn{2}{c||}{$\fNL=0$} & \multicolumn{2}{c||}{$\fNL=50$} & \multicolumn{2}{c||}{$\fNL=100$} & \multicolumn{2}{c}{$\fNL=150$}    \\
\hline
 &$\Delta\fNL$ & $\Delta\nu$ &$\Delta\fNL$ & $\Delta\nu$ & $\Delta\fNL$ &$\Delta\nu$ & $\Delta\fNL$ &$\Delta\nu$  &$\Delta\fNL$ & $\Delta\nu$ &$\Delta\fNL$ & $\Delta\nu$ & $\Delta\fNL$ &$\Delta\nu$ & $\Delta\fNL$ &$\Delta\nu$   \\
\hline
Local & $1.2$& -& $2.3$& -& $1.6$&-&  $1.7$ &- & $1.2$& -& $2.4$& -& $1.7$&-&  $1.8$ &-\\
\hline
$\nu=1.5$ & $2.8$ & - & $11$&$0.05$&$11$&$0.03$&$7.5$&$0.02$ & $2.9$& - & $19$&$0.09$&$20$&$0.05$&$14$&$0.03$   \\
\hline
$\nu=1.0$ & $9.7$ & - & $9.8$&$0.09$&$9.8$&$0.04$&$9.8$&$0.03$ & $26$ & - & $43$&$0.16$&$43$&$0.08$&$43$&$0.05$  \\
\hline
$\nu=0.5$ & $7.8$ & - & $11$&$0.56$&$11$&$0.28$&$11$&$0.19$ & $29$& - & $44$&$0.77$&$43$&$0.38$&$43$&$0.26$ \\
\hline
Equilateral & $9.9$& -& $9.9$& -& $9.9$&-&  $9.8$ & -& $40$& -& $40$& -& $39$&-&  $39$ -&\\
\hline
\hline
\end{tabular}
\caption{\small Marginalized 1-$\sigma$ errors on the parameters $\fNL$ and $\nu$ expected for our two example surveys V1 and V2 for a choice of fiducial values. For a comparison with other works, we include the constraints expected for the local and equilateral models.}
\label{tb:lss}
\end{table}
\begin{figure*}[t]
{\includegraphics[width=0.8\textwidth]{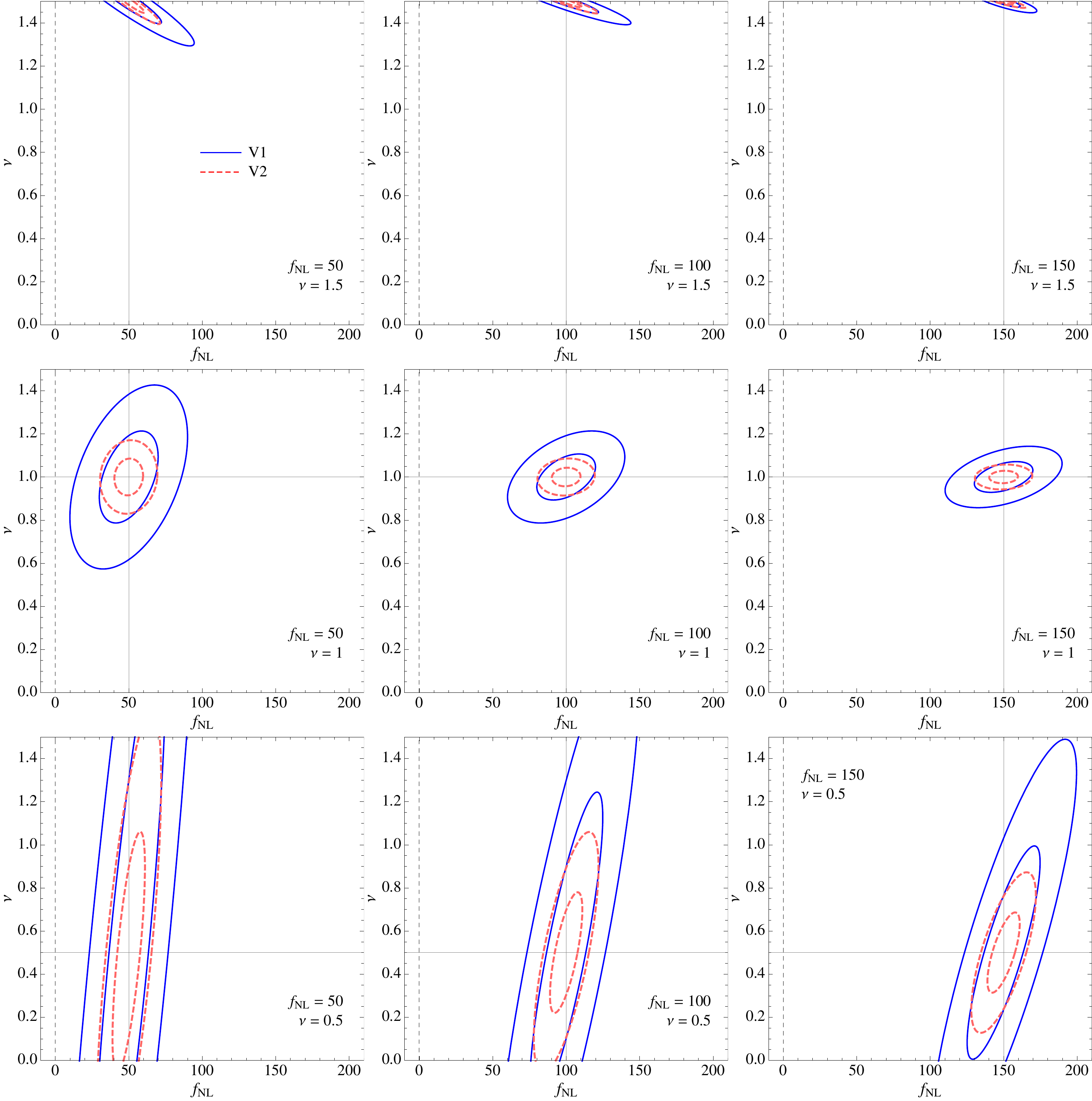}}
\caption{\small 1- and 2-$\sigma$ uncertainty contours corresponding to the determination the parameters $\fNL$ and $\nu$ for three different choices of the fiducial values: $\fNL=50$ ({\em left column}), $100$ ({\em central column}) and $150$ ({\em right column}) and $\nu=1.5$ ({\em top row}), $1.0$ ({\em central row}) and $0.5$ ({\em bottom row}). Continuous blue curves corresponds to the V1 geometry, dashed, red ones to the V2 one. Both cases assume a limiting $k_{\rm max}(z)$ such that $k_{\rm max}(0)=0.15\kMpc$. Caution should be taken when the elliptical contours are large (see discussions in Sec.\,\ref{Sec:CMB_Fisher}.)}
\label{fig:contoursLSS}
\end{figure*}

The marginalized 1-$\sigma$ errors on the $\fNL$ and $\nu$ parameters obtained for the V1 and V2 geometries assuming both $k_{\rm max}=0.15\kMpc$ and $k_{\rm max}=0.075\kMpc$ at $z=0$ are shown in Table~\ref{tb:lss}. In the first place we notice that uncertainties obtained for the local and equilateral parameters are consistent with the results of \citep{GiannantonioEtal2011}. We find in fact lower errors respectively by 30\% and 50\% with respect to their uncertainties when marginalized over cosmological parameters with Planck priors. In the case of the QSF model we observe the expected improvement of the constraints when larger values of both $\fNL$ and $\nu$ are assumed as fiducial values. The dependence on the fiducial values of $\fNL$ and $\nu$ of the 1-$\sigma$ uncertainty $\Delta \nu$ can be very roughly described over the range $0.5\le\nu\le 1.5$, for the V1 geometry, as
\beq
\Delta\nu\simeq\frac{0.1}{\nu^3}\frac{100}{\fNL}\,,
\eeq
with the V2 case corresponding to an uncertainty about a factor of two smaller. The important results is that we can indeed expect to able to distinguish values of $\nu\simeq 1$  from the local limit $\nu=1.5$ for sufficiently large values of $\fNL$, say $\fNL\ge 100$, already with a survey corresponding to our V1 example. Even lower values of $\fNL$ would be sufficient for a larger survey as V2. The improvements in the constraints in the V2 example with respect to V1 are mainly due to the larger volume, although the larger values of the fiducial Gaussian bias and the higher redshift do play a non negligible role.

These considerations are particularly evident in \fig{fig:contoursLSS}, where we show the 1- and 2-$\sigma$ uncertainty contours for $\fNL$ and $\nu$ corresponding to the V1 ({\em blue, continuous curves}) and V2 ({\em red, dashed curves}) examples, assuming the set of fiducial values given by $\fNL=50$, $100$ and $150$ and $\nu=1.5$, $1.0$ and $0.5$. Notice that we choose to keep the same range over the variable $\nu$ for the case of fiducial $\nu=1.5$ as for the other values considered. While such choice makes it hard to distinguish the different curves for large fiducial value of $\fNL$ and $\nu$, it allows, on the other hand, an easier comparison of the results as the fiducial $\nu$ is varied. The complete results of the Fisher analysis are given, in any event, in Table \ref{tb:lss}.

The degeneracy between $\fNL$ and $\nu$ can be easily understood for values of $\nu$ close to the local limit $\nu=1.5$. In this case, in fact, since the dominant effect is given by the scale-dependent correction $\Delta b_{sd}$ and a lower value of $\nu$ can be compensated by an higher $\fNL$. When $\nu$ is close to one, the degeneracy is reduced since the large-scale scale-dependent corrections are balanced by almost scale-independent corrections at small scales. Notice that the bias corrections shown in \fig{fig:db} are evaluated at $z=1$, an at this redshfit the relevant range of scale is given by $0.003\kMpc<k<0.28\kMpc$. At lower values of $\nu$ the error $\Delta \nu$ increases significantly, as the scale-dependent corrections are now very mild and there is a large degeneracy between all the parameters in each redshift bin. The combination of different bins still allows for a relatively low uncertainty on $\fNL$.

\begin{figure*}[t]
{\includegraphics[width=0.8\textwidth]{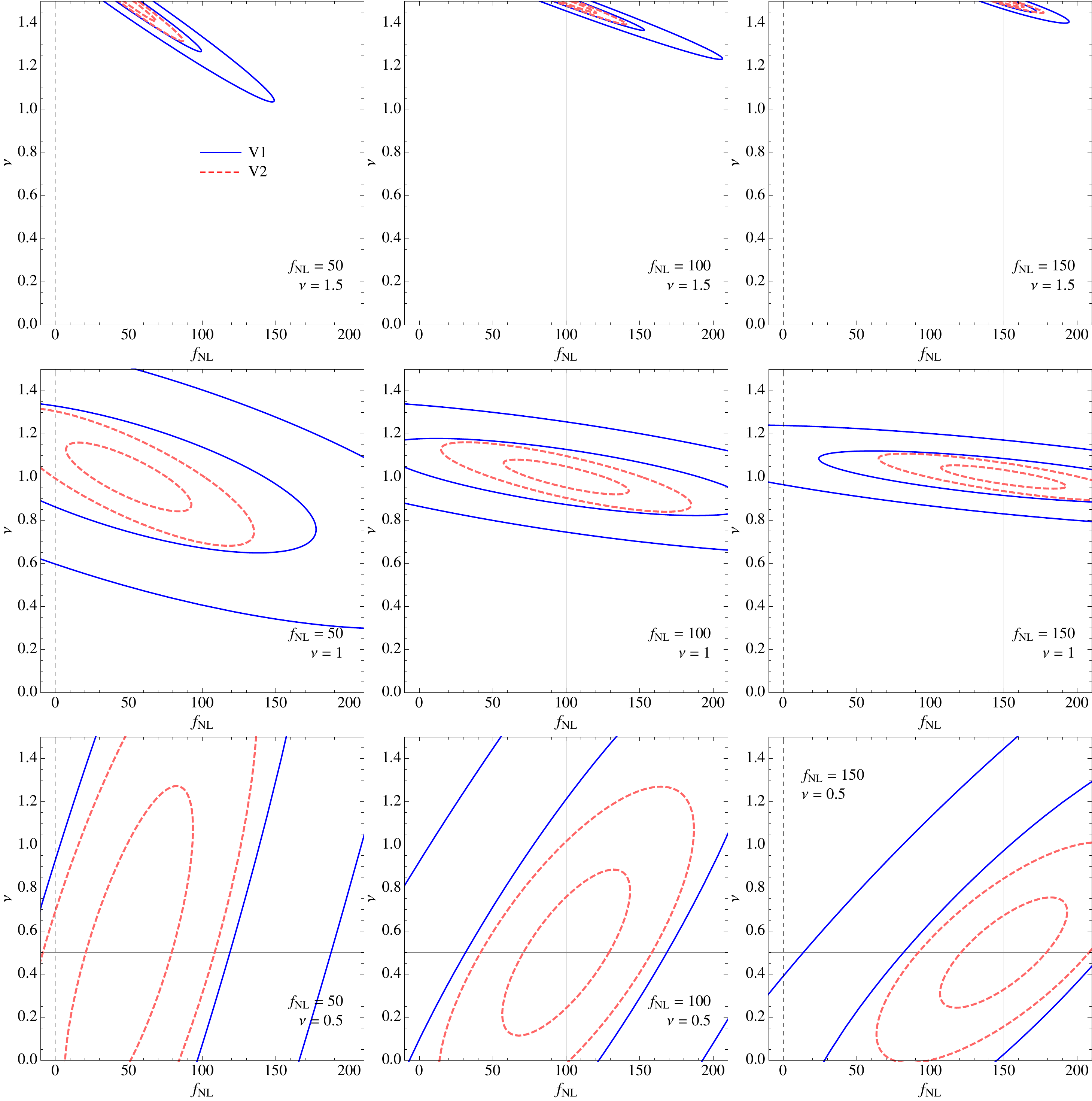}}
\caption{\small Same as \fig{fig:contoursLSS} but for a limiting $k_{\rm max}(z)$ such that $k_{\rm max}(0)=0.075\kMpc$}
\label{fig:contoursLSSB}
\end{figure*}
\fig{fig:contoursLSSB} show the same results as \fig{fig:contoursLSS} but for a limiting $k_{\rm max}(z)$ such that $k_{\rm max}(0)=0.075\kMpc$. We notice, in the first place, a significant worsening of the determination of $\fNL$ already at $\nu=1$. This is due to the lower control over the {\em form} of the power spectrum: the constrains are now mostly based on its {\em amplitude} at large-scales. Remarkably the overall increase in the uncertainty on $\nu$ is only of about a factor of two for all considered fiducial values. The different choice of the range of scales assumed for the analysis results as well in a different degeneracy between $\fNL$ and $\nu$. This is evident for instance for the fiducial values $\fNL=100$ and $\nu=1$. In the conservative case of $k_{\rm max}=0.075\kMpc$ at $z=0$, in fact lower values of $\nu$ are degenerate with higher values of $\fNL$ as expected from the scale-dependence alone. Including smaller scales the degeneracy is greatly reduced and turns slightly in the opposite direction. We should stress the fact that the possibility to properly include in the analysis scales in the mildly nonlinear regime relies on accurate predictions for both the matter power spectrum and bias. Recent developments in the description of matter nonlinearities in Perturbation Theory (see \eg \citep{CrocceScoccimarro2006A, BernardeauCrocceScoccimarro2008, Pietroni2008}) can justify even our more optimistic choice of $k_{\rm max}=0.15\kMpc$ at $z=0$, also in the case of non-Gaussian initial conditions \citep{BartoloEtal2010, BernardeauCrocceSefusatti2010}. A similarly accurate description of halo and galaxy bias in this regime is a more challenging task and will necessarily require further work (see \eg \citep{ChanScoccimarroSheth2012, BaldaufEtal2012} for recent results).

We finally remark that a similar analysis can be extended to the galaxy bispectrum. As shown in \citep{Sefusatti2009, JeongKomatsu2009B, NishimichiEtal2010, GiannantonioPorciani2010, BaldaufSeljakSenatore2011} a scale-dependent correction as the one considered here for linear halo bias is present as well for terms induced by nonlinear, quadratic bias and relevant for the galaxy bispectrum at large scales. A relatively simple model for the halo bispectrum in the presence of local non-Gaussian initial conditions has been recently tested in numerical simulations by \citep{SefusattiCrocceDesjacques2011} where the addition of scale-dependent bias corrections are shown to improve significantly over previous studies of the galaxy bispectrum \citep{ScoccimarroSefusattiZaldarriaga2004, SefusattiKomatsu2007}.

\section{CMB correlations and joint constraints}
\label{sec:cmb}

We now turn to consider the constraints on Quasi-Single Field inflation from the bispectrum of the cosmic microwave sky, notably with the aim of making forecasts relevant for the Planck data.  In principle, given the linearity of the CMB transfer functions,  this is more straightforward than for the matter perturbations.  In practice, however, the task is very computationally intensive so we shall make some simplifying approximations which will yield  constraints and forecasts of reasonable precision.

Our aim is to compare a theoretical prediction for the CMB bispectrum with that obtained from observations, such as WMAP or Planck full sky maps.   The CMB bispectrum is the three-point correlator of the  harmonic coefficients $a_{l m}$ describing the map, $B^{l_1 l_2 l_3}_{m_1 m_2 m_3} = a_{l_1 m_1} a_{l_2 m_2} a_{l_3 m_3}$.
Assuming the bispectrum  has been created by a statistically isotropic process, we can restrict our attention to the angle-averaged bispectrum $\Blll$,
\beq
B_{l_1 l_2 l_3} =  \sum_{m_i}h_{l_1 l_2 l_3}^{-1}  \curl{G}^{l_1 l_2 l_3}_{m_1 m_2 m_3} a_{l_1 m_1} a_{l_2 m_2} a_{l_3 m_3}\,,
\eeq
where $h_{l_1 l_2 l_3}$ is a geometrical factor which enforces a multipole triangle condition,
\beq
h_{l_1 l_2 l_3} =  \sqrt{\frac{(2l_1+1)(2l_2+1)(2l_3+1)}{4\pi}} \( \begin{array}{ccc} l_1 & l_2 & l_3 \\ 0 & 0 & 0 \end{array} \)\,,
\eeq
and $ \curl{G}^{\,\,l_1\; l_2\; l_3}_{m_1 m_2 m_3}$ is the Gaunt integral,
\begin{align}\label{eq:Gaunt}
 \curl{G}^{l_1 l_2 l_3}_{m_1 m_2 m_3} =h_{l_1 l_2 l_3} \( \begin{array}{ccc} l_1 & l_2 & l_3 \\ m_1 & m_2 & m_3 \end{array} \)\,,
\end{align}
with  the usual Wigner-$3j$ symbol.    It is generally more straightforward to work with the reduced bispectrum $b_{l_1 l_2 l_3} = h_{l_1 l_2 l_3}^{-1} B_{l_1 l_2 l_3}$ with the geometrical factors removed.

To find the reduced CMB bispectrum $b_{l_1 l_2 l_3}$ induced by a given primordial bispectrum $B_\O(k_1,k_2,k_3)$, we use the CMB transfer functions $\D_{l}(k)$  to project forward as
\begin{align}
\label{eq:redbispect}
\nn b_{l_1 l_2 l_3}= \(\frac{2}{\pi}\)^3 \int & dx d k_1 d k_2 d k_3\, \(x k_1 k_2 k_3\)^2\, B_\O(k_1,k_2,k_3)\\
& \D_{l_1}(k_1) \D_{l_2}(k_2) \D_{l_3}(k_3)\, j_{l_1}(k_1 x) j_{l_2}(k_2 x) j_{l_3}(k_3 x)\,.
\end{align}
If the original  bispectrum $B_\Phi(k_1,k_2,k_3)$ is separable, the complicated 4D integral (\ref{eq:redbispect}) also separates and becomes much more tractable.   While the local model and the usual equilateral ansatz are separable, the quasi-local shape, \eqn{eq:tempQsA}, which interpolates between them is  is not separable.   Nevertheless, for a given QSF parameter $\nu$, we can still calculate $b_{l_1 l_2 l_3}$  by using a separable eigenmode expansion, but we will not describe the modal methodology here in detail (see \citep{FergussonLiguoriShellard2010B}).

\subsection{CMB bispectrum correlator}

In order to determine if a given theoretical bispectrum is present in the observational data, we employ the approximate estimator
\begin{align}\label{eq:approxestimator}
\curl{E} = \frac{1}{\tilde{N}^2} \sum_{l_i m_i} \frac{\curl{G}^{l_1 l_2 l_3}_{m_1 m_2 m_3} \, {b}_{l_1 l_2 l_3} }{{C}_{l_1}{C}_{l_2}{C}_{l_3} } a_{l_1 m_1} a_{l_2 m_2} a_{l_3 m_3}\,,
\end{align}
with appropriate experimental effects incorporated - beam, noise and mask.  The estimator essentially performs a least squares fit between theory and data with the ratio above yielding the signal-to-noise.  We can, in principle, use the separable modal methodology in ref.~\citep{FergussonLiguoriShellard2010B} to find constraints on quasi-single field inflation.  The reconstructed bispectrum coefficients extracted from the WMAP7 data  \citep{FergussonLiguoriShellard2010B} can be used to compare directly with the expansion coefficients predicted theoretically from \eqn{eq:redbispect}.

\begin{figure}[t]
\centering
\includegraphics[width=.4\textwidth]{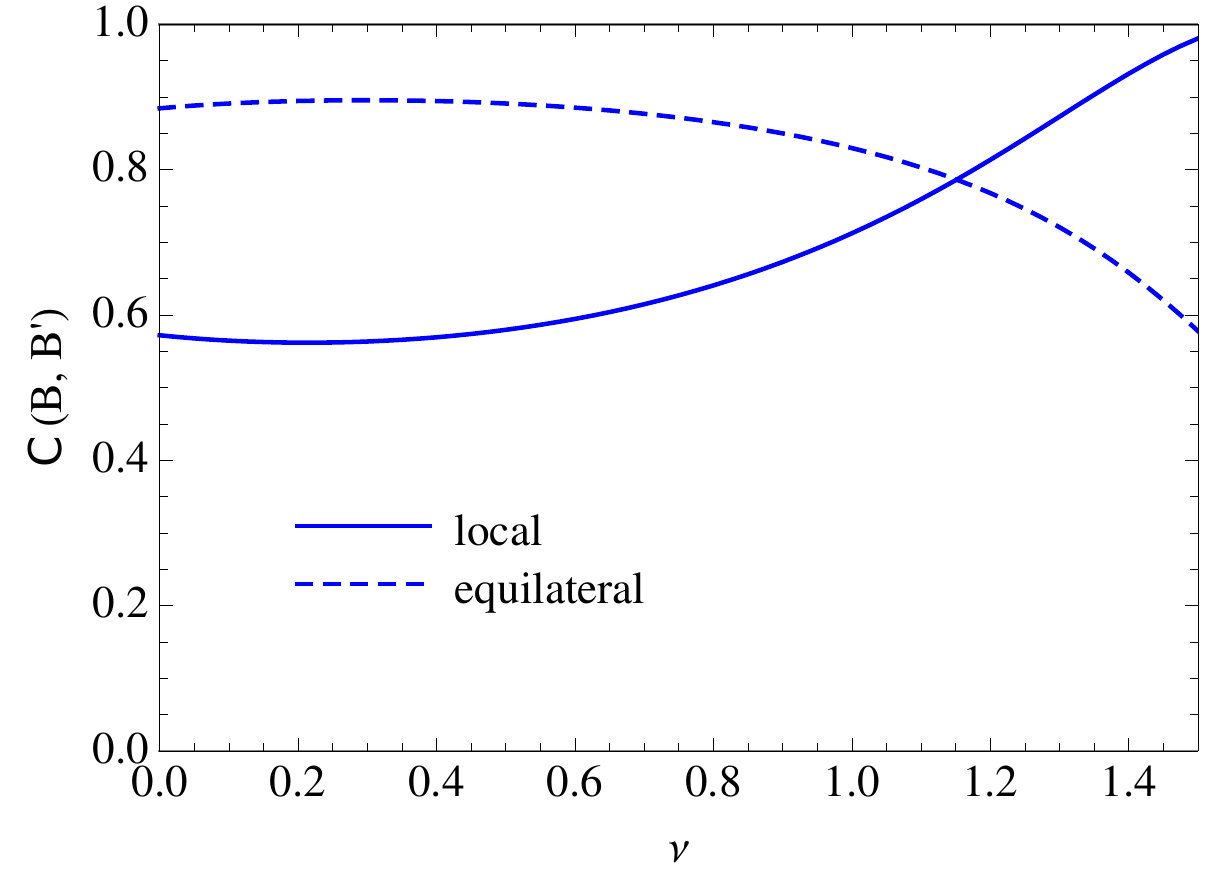}
\includegraphics[width=.5\textwidth]{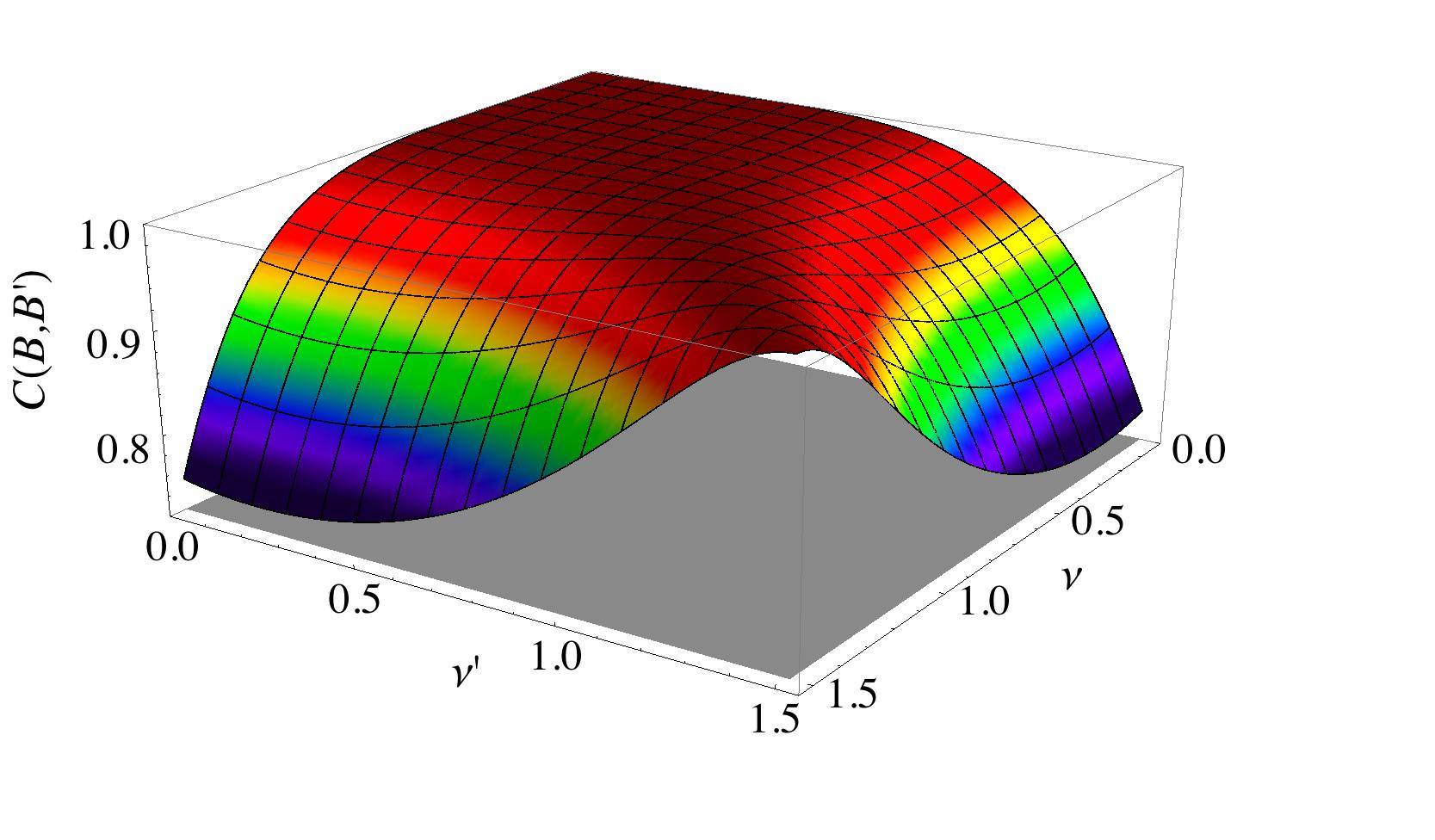}
\caption[]{\small {\em Left panel}: Estimate of the CMB correlation of the quasi-single field  shape as a function of the parameter $\nu$ with the local and equilateral shapes using the weighted shape correlator \eqn{eq:shapecor}.   Note that the  $\nu = 1.5$ model is nearly perfectly correlated with the local shape, whereas an 88\% correlation is achieved in the equilateral limit $\nu\rightarrow 0$. {\em Right panel}: Estimate of the CMB correlations of the QSF models with themselves ${\cal C}(B,B')$ as a function of the parameters $\nu,\, \nu'$. Note the large plateau of strong correlations for models with $0\le \nu \le 0.75$ and $0\le \nu' \le 0.75$, which makes it difficult distinguish QSF models with small $\nu$.}
\label{fig:cmb_corr}
\end{figure}

Whether or not two values of the quasi-single field bispectrum parameter $\nu$ can be distinguished by a given data set can be determined by a Fisher matrix analysis, essentially the cross-correlator between the two CMB bispectra
\begin{align}\label{eq:cmbcor}
\curl{C}(B,B^\pr) = \frac{1}{N}\sum_{l_i} \frac{B_{l_1 l_2 l_3}B^\pr_{l_1 l_2 l_3}}{C_{l_1} C_{l_2} C_{l_3}}\,,
\end{align}
where the normalization is defined by
\begin{align}
N = \sqrt{ \sum_{l_i}\frac{B^2_{l_1 l_2 l_3}}{C_{l_1} C_{l_2} C_{l_3}}} \sqrt{\sum_{l_i} \frac{{B^\pr}^2_{l_1 l_2 l_3}}{C_{l_1} C_{l_2} C_{l_3}}}\,.
\end{align}
However, this approach is extremely computationally demanding as we must calculate the full bispectrum for each value of $\nu$ before we can make any comparison. This has been achieved already for local and equilateral asymptotes but we would like a simpler method for estimating the interpolants.   As shown in \citep{FergussonShellard2009}, a fairly accurate measure of the Fisher matrix, \eqn{eq:cmbcor}, can be obtained from the shape correlator,
\begin{align}\label{eq:shapecor}
\bar{\curl{C}}(S,S^\pr) = \frac{F(S,S^\pr)}{\sqrt{F(S,S)F(S^\pr,S^\pr)}}\,.
\end{align}
where
\begin{align}\label{eq:shapeint}
F(S,S^\pr) =  \int_{\curl{V}_k} S(k_1,k_2,k_3) \,S^\pr(k_1,k_2,k_3) \,\w(k_1,k_2,k_3) \, d\curl{V}_k\,,
\end{align}
with the bispectrum `shape' $S(k_1,k_2,k_3)$ defined from the rescaled primordial bispectrum
 \begin{align} \label{eq:shapefn}
S(k_1,k_2,k_3) = (k_1 k_2 k_3)^2 B_\Phi(k_1,k_2,k_3)\,.
\end{align}
Here, in order to replicate the scaling in the CMB bispectrum correlator, \eqn{eq:cmbcor}, we adopt the weight function in \eqn{eq:shapeint} as
\begin{align}\label{eq:weight}
w(k_1,k_2,k_3) = \frac{1}{k_1+k_2+k_3}\,.
\end{align}
It is demonstrated in ref.~\citep{FergussonShellard2009} that \eqn{eq:shapecor} yields a good phenomenological approximation to \eqn{eq:cmbcor} for a wide variety of shapes (all classes of scale-invariant models). For example, the shape cross-correlator $\curl{C}(S,S^\pr)$  for the local and equilateral models, produced only a 5\% underestimate of the true CMB correlator $\curl{C}(B,B^\pr)$ at Planck resolution.  This essentially sets an upper limit on the accuracy of the weighted shape correlator for estimating the CMB correlations between the local, equilateral and QSF models.   It is sufficient precision for making Planck forecasts in the present context.

Estimates of CMB correlations for the QSF models using the shape correlator of \eqn{eq:shapecor} are illustrated in \fig{fig:cmb_corr} ({\em left panel}). It is clear from \fig{fig:cmb_corr} that the QSF model with $\nu=1.5$ is very well approximated by the local shape, while smaller values across a broad range $\nu \le 0.75$ are close to equilateral, though they never attain more than a 90\% correlation.   The two-dimensional cross-correlation for the QSF models with themselves (\ie as functions of $\nu$, $\nu'$) from \fig{fig:cmb_corr} ({\em right panel}) indicate that it will be very difficult to distinguish between models with $\nu,\nu' <0.75$.   However, there are significant differences between models with small  $ \nu < 0.75$ and large $\nu'>1.0$.

\begin{figure}[t]
\centering
\includegraphics[width=.9\linewidth]{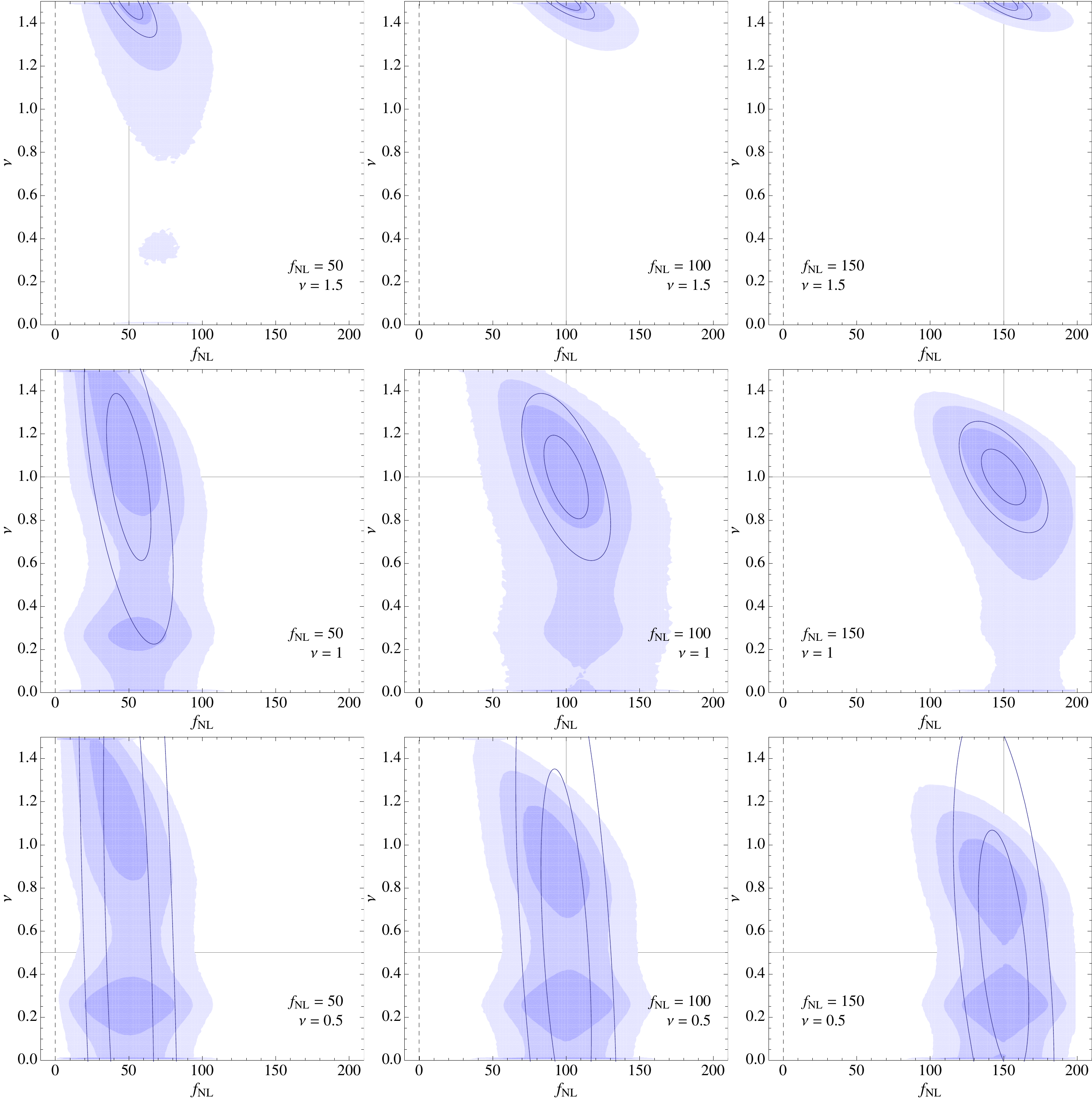}
\caption[]{\small Expected joint constraints on the two parameters $\fNL$ and $\nu$ from measurements of temperature bispectrum from Planck-like CMB observations. The shaded areas of decreasing colour density corresponds to the 1-, 2- and 3-$\sigma$ constraints from the forecasted likelihood function. For comparison, the Fisher matrix results corresponding to 1- and 2-$\sigma$ constraints are show by superimposed ellipses.  These are qualitatively in agreement and identify the correct degeneracy directions, but become inaccurate in degenerate regions for small $\nu$.}
\label{fig:cmb_contours}
\end{figure}

\subsection{CMB Fisher matrix and likelihood analysis}
\label{Sec:CMB_Fisher}

\begin{figure}[t]
\centering
\includegraphics[width=.5\linewidth]{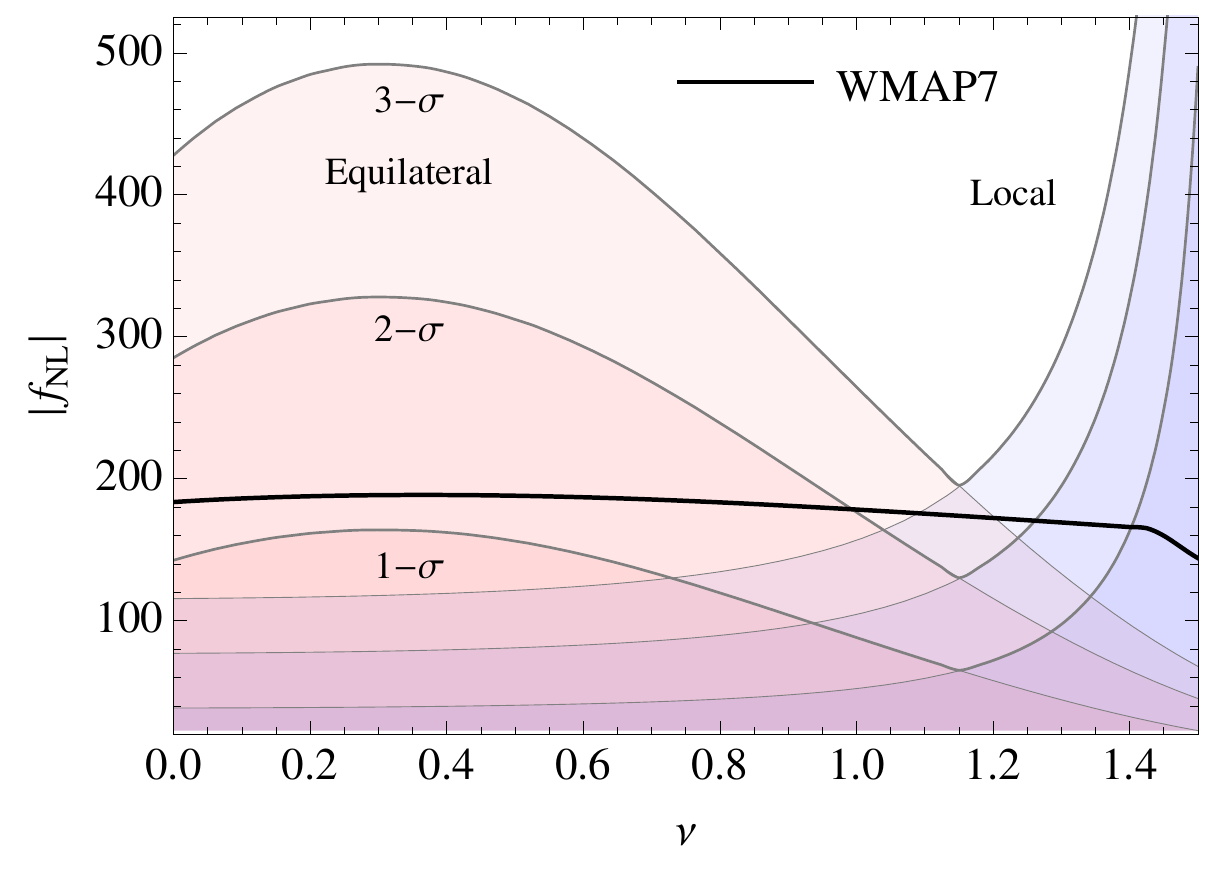}
\includegraphics[width=.37\linewidth]{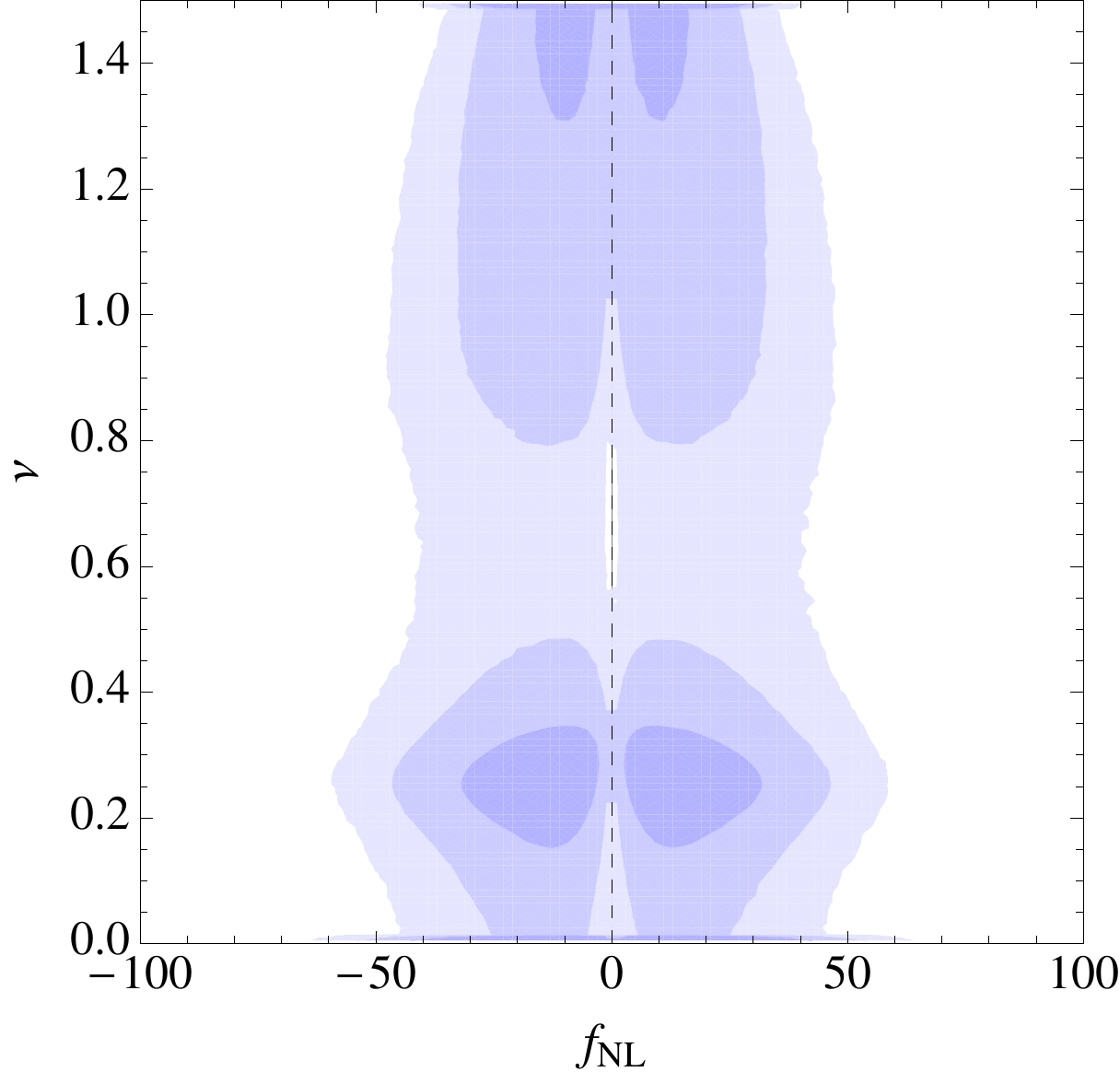}
\caption[]{\small {\em Left panel}: Minimal values of $f_{\rm NL}$ required to separate the QSF shape from either the spurious local or spurious equilateral signal
it would generate at different significance levels. The black continuous curve shows the value of $\fNL$ for which either the spurious local or spurious equilateral signal would violate the WMAP7 2-$\sigma$ bounds on these parameters. {\em Right panel}: Expected joint constraints on the two parameters $\fNL$ and $\nu$ from measurements of temperature bispectrum from Planck-like CMB observations, assuming the fiducial value $\fNL=0$. The shaded area of decreasing colour density correspond to the 1-, 2- and 3-$\sigma$ constraints from the forecasted likelihood function.  Note that volume factors in the transformation from the orthonormal `observation' space back to the $\{\fNL,\,\nu\}$-space yields a non-uniform distribution for $\nu$ which disfavors values of $\nu\approx 0.7$.}
\label{fig:cmb_constraint}
\end{figure}

In order to forecast the implications of the Planck experiment for QSF models, we have performed a two-dimensional CMB Fisher matrix analysis analogous to that for the large-scale structure in Section\,\ref{sec:fisher}, as well as a likelihood analysis.  Here we have defined a CMB Fisher matrix ${\cal F}_{a,b}$ for the parameters $ p_a = \{\fNL,\,\nu\}$ as in \eqn{eq:fisherLSS} using knowledge of the approximate correlator, \eqn{eq:shapecor}, for the QSF shape functions.   We do not marginalize over cosmological parameters for which we believe there is only a weak dependence for the CMB bispectrum in any case.   Assuming that Planck achieves a target 1-$\sigma$ variance $\Delta \fNLl =5$, in Figure\,\ref{fig:cmb_contours} we have plotted 1-$\sigma$ and 2-$\sigma$ uncertainty contours for $\fNL$ and $\nu$ for the same fiducial values as previously: $\fNL = 50$, $100$ and $150$ and $\nu= 1.5$, $1.0$ and $0.5$.   These CMB Fisher matrix ellipses ({\em continuous curves}) have degeneracy directions which are complementary to the large-scale structure analysis, and so there is value in a joint analysis as we shall discuss.   These results show that  $\nu \approx 1.0$ models will be distinguishable from both equilateral or local models at better than the 2-$\sigma$ level for $\fNL\ge 100$ (which is approximately equivalent when normalised relative to the local model to $\fNLl\ge 30$).   Nevertheless the results also show the deficiency of the Fisher matrix ellipses for models near $\nu,\nu' \le 0.75$ where the correlation functions are very flat or degenerate.   The large elliptical contours obtained from the local derivatives around $\nu=0.5$ appear to indicate that these models cannot be distinguished from  the local  models, which is apparently not consistent with the $\nu\ge 1.0$ results at large $\fNL$.

The shortcomings of this simple Fisher matrix analysis motivated an improved likelihood analysis based on the approximate correlator, \eqn{eq:shapecor}.   This was achieved in several steps. First, the Fisher matrix ${\cal F}$ for $n$$=$$300$ values of $\nu$ (denote these discrete values $\nu_j$) was calculated on a uniform $n^2$ grid using \eqn{eq:shapeint}.  All matrix elements were divided by the local-local correlation result, so that the diagonal elements of this matrix yield the QSF model normalization at a given $\nu_j$ relative to the local bispectrum.    Of course, when normalised with \eqn{eq:shapecor} then ${\cal F}$ yields the correlation matrix ${\cal C}$ illustrated in Fig.\,\ref{fig:cmb_corr} ({\em right panel}). Secondly, we diagonalize the Fisher matrix to extract the optimal uncorrelated basis, that is, representing the matrix ${\cal F}_{ij}$ as
\begin{align}
{\cal F} = {\cal V}\,{\cal D}\, {\cal V}^T\,
\end{align}
where ${\cal V}$ contains the unit eigenvectors $v_i$ and ${\cal D}$ is a diagonal matrix constructed from an ordered list of the eigenvalues $\lambda_i$.   PCA analysis of ${\cal F}$ for these QSF models reveals that just two eigenvalues are dominant, with eigenvectors representing the local shape and an equilateral-like shape.  A third eigenvector makes a relatively small additional contribution for a more accurate analysis.

We next perform Monte Carlo simulations to populate data distributions about a specific $\{\fNL,\,\nu_j\}$ model which are consistent with the Planck variance $\Delta \fNLl = 5$.    To achieve this, we need to know how much of each eigenvector $v_i$ is required to build a particular correlation curve, corresponding to the given $\nu_j$ value.    The corresponding data `shape' vector $\textstyle \alpha_i^{{\nu_j}}$ for the $\nu_j$ model can be found as the $j^{\rm th}$ row of the matrix
\begin{align}
\alpha_i^{{\nu_j}}=( {\cal V}\sqrt{\cal D})_{ij}\,,
\end{align}
which is automatically unit normalised.  We then multiply by the required $\fNL$ (again normalised to the local model) and the appropriate noise vector $N_j$ is added, which consists of $n$ Gaussian random variables of variance $5$, that is, the data vector becomes
\begin{align}
d_i =\fNL \,\alpha_i^{{\nu_j}} + N_i\,.
\end{align}
We can do this essentially because we are working in an orthonormal `observation' space in which the probability distribution is a multivariate Gaussian (this understanding of the independent variance in orthogonal modes is exploited, for example, in modal CMB map simulations described in ref.~\citep{FergussonLiguoriShellard2010B}). We must then map back to the $\{\fNL,\,\nu\}$ space in which $\fNL$ (normalised to the local model) for each given original value of $\nu_j$ is given by
$\fNL =  \sum_i d_i \alpha_i^{{\nu_j}}$ and the best fit $\nu$ for this data is the one that minimises the error $ \sum_i (d_i - \fNL \,\alpha_i^{{\nu_j}})^2$.   This process is repeated many times to build up a smooth likelihood function about the given model.   This mapping from the orthonormal `observation' space back to $\{\fNL,\,\nu\}$ space does not necessarily recover a peak in the  distribution around the original model values because the Jacobian for the transformation rescales the volume factor.   The non-trivial weighting in the $\{\fNL,\,\nu\}$ `theory' space introduces artifacts like those shown in Fig.\,\ref{fig:cmb_constraint} ({\em right panel}) where $\nu$ does not even have a uniform probability distribution about the null $\fNL =0$ model.   This indicates that it can be useful to choose parametrisations for theoretical models which match, as closely as possible, the signal-to-noise weighting mandated by the observational data.

The results of the CMB likelihood analysis are illustrated in Fig.\,\ref{fig:cmb_contours} with 1-$\sigma$, 2-$\sigma$ and 3-$\sigma$ contour shading.  A direct comparison to the Fisher matrix ellipses is made for the same input model values as above.     While the results are in qualitative agreement and identify the same degeneracy directions, they also illustrate the limitations of a Fisher matrix analysis which requires a `global' extrapolation into regions of the parameter space where the distributions are either rapidly changing or degenerate, neither of which can be adequately captured by local derivatives.  Additional features appear in the likelihood analysis, particularly note the two peaks  in the nearly degenerate $\nu=0.5$ case.   For small $\nu$,  the likelihood contours are cut off before reaching the local model  ($\nu=1.5$), so there are parameter values for the QSF model which can be distinguished from both local and equilateral shapes for $\fNL^{\rm QSF} \ge 100$.

In \fig{fig:cmb_constraint} ({\em left panel}) we illustrate the expected $f_{\rm NL}$ signal required to distinguish QSF from the local and equilateral models at various significance levels.  We also plot the present WMAP7 constraints on any QSF model from the union of local model and equilateral model estimation results \citep{KomatsuEtal2011} (assuming the correlations shown in  \fig{fig:cmb_corr}).  As can be seen, it will be difficult  to distinguish equilateral from QSF models with $\nu<1.0$ where the correlation function ${\cal C}$ is very flat;  a large measurement $f_{\rm NL}>200$ is required and this is excluded already by WMAP7.   Similarly, it will not be possible to distinguish the local model from QSF models in the narrow region above $\nu > 1.3$ because the necessary $f_{\rm NL}$ is also excluded.   This leaves a narrow window of opportunity for Planck data to identify QSF models for $1.0 <\nu < 1.3$ assuming a large $f_{\rm NL} > 120$ measurement  (approximately equivalent to a local $\fNL^{\rm loc.} \ge 40$).   Using Planck polarization data with a reduced variance  $\Delta f_{\rm NL} \approx 3$, the forecast window widens considerably to  $0.5 <\nu < 1.4$ and offers a much better prospect of differentiating between the QSF and other models at greater than a 2$\sigma$ significance.

\subsection{Combined CMB and LSS forecasts}
\label{sec:combined}

In this section we consider the combination of our results for the LSS Fisher matrix analysis with those from the approximate CMB analysis. This combination is particularly interesting because, as mentioned above, the CMB bispectrum and the galaxy power spectrum observations weight differently distinct triangular configurations of the initial bispectrum. For this reason we expect that a joint forecast should reduce the specific degeneracies between the two parameters $\fNL$ and $\nu$ already highlighted in the previous sections.

\begin{figure*}[t]
{\includegraphics[width=0.8\textwidth]{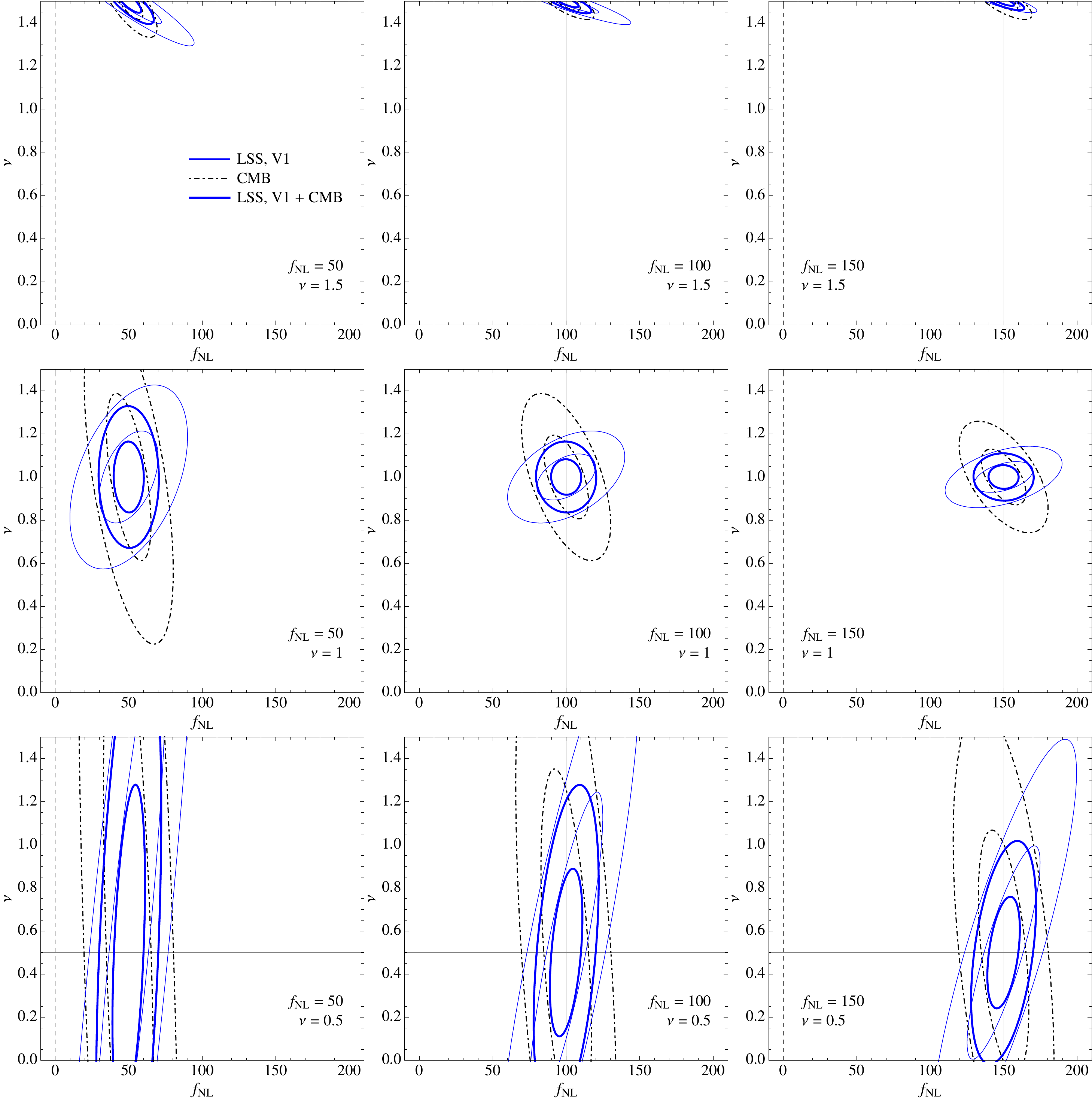}}
\caption{\small 1- and 2-$\sigma$ CMB contours ({\em black, dot-dashed curves}), LSS contours corresponding to the V1 survey, assuming $k_{\rm max}=0.15\kMpc$ at $z=0$ ({\em blue, continuous, thin curves}) and the contours corresponding to the joint CMB and LSS analysis ({\em blue, continuous, thick curves}). As in the previous plots we consider the set of fiducial values given by all combinations of $\fNL=50$, 100 and 150 and $\nu=0.5$, 1 and $1.5$.}
\label{fig:contoursLSSCMBa}
\end{figure*}

\begin{figure*}[t]
{\includegraphics[width=0.8\textwidth]{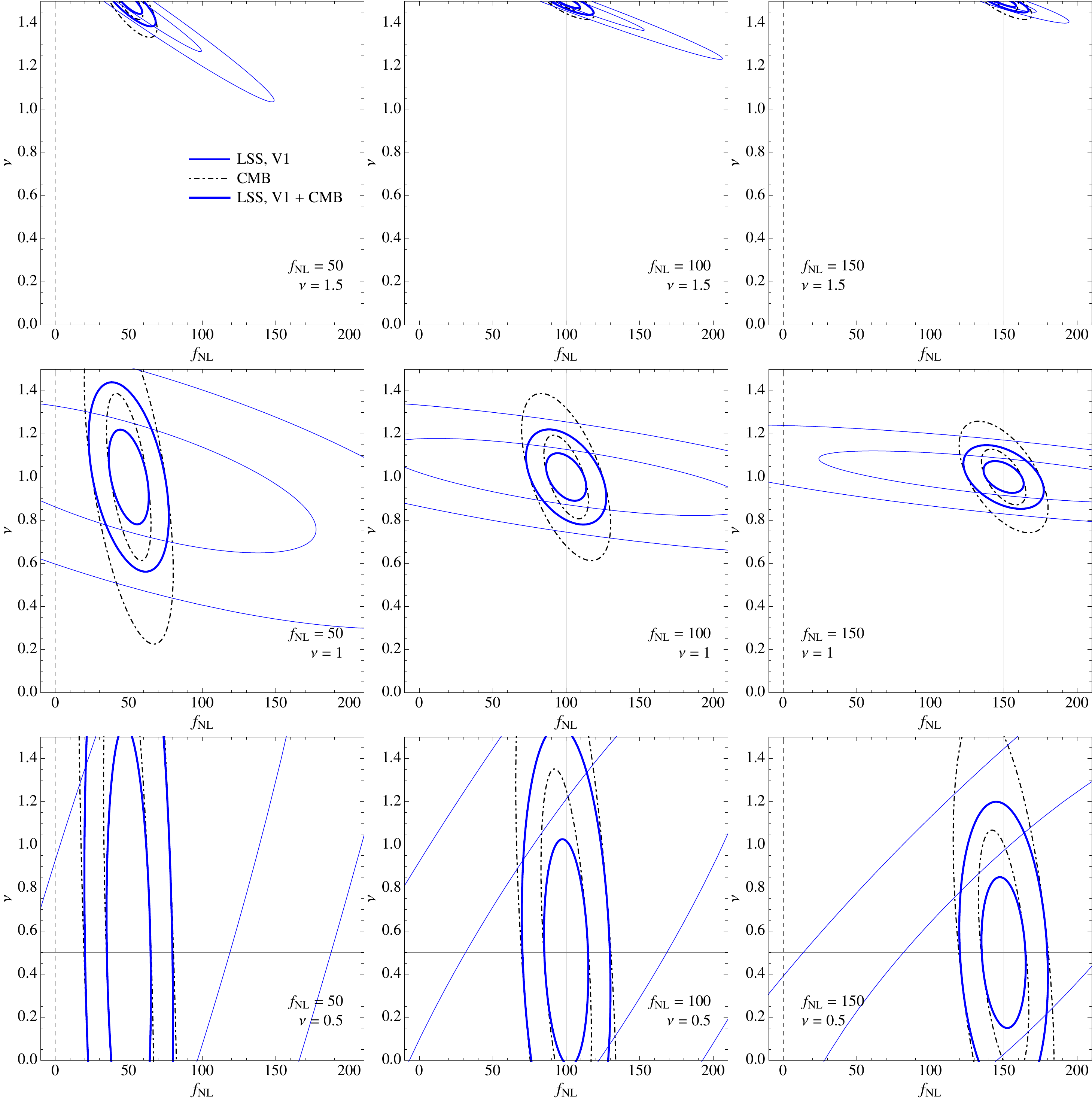}}
\caption{\small Same as Figure~\ref{fig:contoursLSSCMBa} but with the more conservative choice of $k_{\rm max}=0.075\kMpc$ at $z=0$ for the galaxy power spectrum analysis.}
\label{fig:contoursLSSCMBb}
\end{figure*}

Since we did not consider a full likelihood analysis for the LSS forecasts, we will limit ourselves to comparing the Fisher matrix results. \fig{fig:contoursLSSCMBa} shows the (1- and 2-$\sigma$) CMB contours, the LSS contours corresponding to the V1 survey, assuming $k_{\rm max}=0.15\kMpc$ at $z=0$, and the contours corresponding to the joint CMB and LSS analysis. The complementarity of the two observables is evident, for instance, in the case given by the fiducial values $\fNL=100$ and $\nu=1$ ({\em central panel}). We remark, again, that the large-scale structure results rely on our ability to accurately describe non-Gaussian correction to galaxy bias at mildly nonlinear scales, where the scale-independent component of such corrections dominates. This can be seen, for instance, from the particular degeneracy between $\fNL$ and $\nu$. As already mentioned, an accurate theoretical model for galaxy bias at these scales will require further investigations and comparison with numerical simulations.

The same joint results, but assuming the more conservative choice of $k_{\rm max}=0.075\kMpc$ at $z=0$ for the galaxy power spectrum analysis, are shown in \fig{fig:contoursLSSCMBb}. In this case the scale-dependent correction to bias represent the dominant effect of primordial non-Gaussianity on the LSS observable. Remarkably, for a fiducial $\nu= 1$, while the uncertainty on $\fNL$ from the galaxy power spectrum is very large, the addition of LSS information can significantly improve the CMB constraints on $\nu$.

\begin{table}[t]
\centering
\begin{tabular}[t]{l||c|c||c|c||c|c||c|c||c|c||c|c}
\hline
\hline
  \multicolumn{13}{l}{Planck}\\
  \hline
  & \multicolumn{2}{c||}{$\fNL=50$} & \multicolumn{2}{c||}{$\fNL=100$} & \multicolumn{2}{c||}{$\fNL=150$}  & \multicolumn{6}{c}{}    \\
\hline
 &$\Delta\fNL$ & $\Delta\nu$ &$\Delta\fNL$ & $\Delta\nu$ & $\Delta\fNL$ &$\Delta\nu$ & \multicolumn{6}{c}{}  \\
\hline
$\nu=1.5$  & $9.6$&$0.08$&$9.6$&$0.04$&$9.6$&$0.04$  & \multicolumn{6}{c}{}  \\
\hline
$\nu=1.0$ & $15$&$0.39$&$15$&$0.19$&$15$&$0.13$ & \multicolumn{6}{c}{}    \\
\hline
$\nu=0.5$  & $17$&$1.7$&$17$&$0.85$&$17$&$0.57$  & \multicolumn{6}{c}{}  \\
\hline
\hline
  \multicolumn{13}{l}{Planck + V1 survey ($V=103\cGpc$, $0.5<z<2$)}\\
  \hline
 &  \multicolumn{6}{c||}{$k_{\rm max}=0.15\kMpc$ at $z=0$} & \multicolumn{6}{c}{$k_{\rm max}=0.075\kMpc$ at $z=0$}\\
  \hline
 &  \multicolumn{2}{c||}{$\fNL=50$} & \multicolumn{2}{c||}{$\fNL=100$} & \multicolumn{2}{c||}{$\fNL=150$}  & \multicolumn{2}{c||}{$\fNL=50$} & \multicolumn{2}{c||}{$\fNL=100$} & \multicolumn{2}{c}{$\fNL=150$}    \\
\hline
 & $\Delta\fNL$ &$\Delta\nu$ & $\Delta\fNL$ &$\Delta\nu$  &$\Delta\fNL$ & $\Delta\nu$ &$\Delta\fNL$ & $\Delta\nu$ & $\Delta\fNL$ &$\Delta\nu$ & $\Delta\fNL$ &$\Delta\nu$   \\
\hline
$\nu=1.5$ & $8.5$&$0.05$&$8.6$&$0.03$&$8.6$&$0.03$ &    $9.2$&$0.06$&$9.3$&$0.03$&$9.3$&$0.03$   \\
\hline
$\nu=1.0$ & $10$&$0.16$&$10$&$0.08$&$10$&$0.07$&    $14$&$0.22$&$14$&$0.11$&$14$&$0.09$   \\
\hline
$\nu=0.5$ & $11$&$0.78$&$11$&$0.39$&$11$&$0.33$ &   $15$&$1.0$&$15$&$0.53$&$15$&$0.43$ \\
\hline
\hline
  \multicolumn{13}{l}{Planck + V2 survey ($V=390\cGpc$, $0.4<z<3.6$)}\\
  \hline
 &  \multicolumn{6}{c||}{$k_{\rm max}=0.15\kMpc$ at $z=0$} & \multicolumn{6}{c}{$k_{\rm max}=0.075\kMpc$ at $z=0$}\\
  \hline
 &  \multicolumn{2}{c||}{$\fNL=50$} & \multicolumn{2}{c||}{$\fNL=100$} & \multicolumn{2}{c||}{$\fNL=150$}  & \multicolumn{2}{c||}{$\fNL=50$} & \multicolumn{2}{c||}{$\fNL=100$} & \multicolumn{2}{c}{$\fNL=150$}    \\
\hline
 & $\Delta\fNL$ &$\Delta\nu$ & $\Delta\fNL$ &$\Delta\nu$  &$\Delta\fNL$ & $\Delta\nu$ &$\Delta\fNL$ & $\Delta\nu$ & $\Delta\fNL$ &$\Delta\nu$ & $\Delta\fNL$ &$\Delta\nu$   \\
\hline
$\nu=1.5$ &  $7.0$&$0.04$&$7.1$&$0.02$&$7.1$&$0.02$  &    $8.3$&$0.04$&$8.6$&$0.02$&$8.6$&$0.02$   \\
\hline
$\nu=1.0$ & $7.7$&$0.08$&$7.7$&$0.04$&$7.7$&$0.04$ &    $13$&$0.11$&$13$&$0.05$&$13$&$0.05$  \\
\hline
$\nu=0.5$ & $8.4$&$0.50$&$8.3$&$0.23$&$8.1$&$0.21$  &   $14$&$0.50$&$14$&$0.25$&$14$&$0.23$  \\
\hline
\hline
\end{tabular}
\caption{\small Marginalized 1-$\sigma$ errors on the parameters $\fNL$ and $\nu$ expected from the Fisher matrix analysis of the Planck temperature bispectrum alone and in combination with the two example LSS surveys V1 and V2 for a choice of fiducial values.}
\label{tb:cmblss}
\end{table}
Table~\ref{tb:cmblss} shows the marginalized 1-$\sigma$ errors predicted by the CMB Fisher matrix analysis. In addition, also shown is the combination of the CMB analysis with the LSS results for the V1 and V2 surveys assuming  the two choices for the range of scales included considered in Section~\ref{sec:halobias}.

\section{Conclusions}
\label{sec:conclusions}

In this work we have studied the effects of the primordial non-Gaussianity predicted by Quasi-Single Field models of inflation on large-scale structure observables with particular attention to the galaxy power spectrum and to the induced scale-dependent correction to halo and galaxy bias. In addition we studied the detectability and the correlation with other non-Gaussian models in CMB bispectrum measurements.

Scale-dependent bias corrections have been the subject of several studies in recent years as they provide a remarkable test of squeezed configurations of the initial bispectrum. In the particular case of the local model, where the primordial curvature bispectrum peaks precisely in the squeezed limit, constraints on the amplitude parameter $\fNL$ from current measurements of the power spectrum of high-redshift sources are already comparable to CMB results. In this context, Quasi-Single Field models represent a veritable case study as they predict a one-parameter family of curvature bispectra with variable momentum-dependence in the squeezed limit, resulting in a correction to linear bias with a scale-dependence interpolating between the one predicted by the local model to practically scale-independence.

In addition to this interesting phenomenological aspect, the determination of the parameter $\nu$ can also provide direct information on the high-energy theory. Supersymmetry naturally determines the range of masses of light scalars during inflation. Interestingly, such masses may be directly observable through QSF inflation models, in terms of the powers in the peculiar scale-dependence in galaxy bias and/or momentum-dependence in CMB non-Gaussianities. In this paper we investigated how much, in the event of the discovery of large primordial non-Gaussianities, we can refine such discovery and tell the difference between QSF predictions from those of the local or equilateral shapes in the future LSS and CMB experiments.

We find that there is an interesting cosmic complementarity between the galaxy halo bias and CMB in the two-parameter space, $\nu$ and $\fNL$. The halo bias is relatively more sensitive to $\nu$ since the scale-dependent correction mostly depend on the squeezed-limit behavior of the primordial bispectrum. The CMB is more sensitive to $\fNL$ since it has more signal weight away from the squeezed limit. By combining both, depending on the value of $\fNL$, a significant fraction of the parameter space in $\nu$ may be distinguished from the local model, although for $\nu$ very close to $3/2$ the degeneracy will remain.

In this work we have concentrated on the primordial bispectra. Nonetheless QSF models also predict large trispectra with similar peculiar momentum dependence. Furthermore, at least in the perturbative region, these models predict that the size of the trispectra $t_{NL}^{\rm SE}$ (from the scalar-exchange diagram) is much larger than the size of the bispectra squared $\fNL^2$ \cite{ChenWang2010B}. In particular, various soft momentum limits of the trispectra have corresponding intermediate momentum dependence determined by the isocurvaton mass \cite{AssassiBaumannGreen2012}. These make the observability of trispectra more prominent comparing to models with $t_{NL}^{\rm SE} \sim \fNL^2$ \cite{ChenEtal2009, ArrojaEtal2009, FergussonReganShellard2010}. It remains to be seen, however,  whether the promise offered by the trispectra can contribute further towards the goal of this paper.\\

During the completion of this project we became aware of a similar study in preparation \cite{NorenaEtal2012}. A rough comparison with their results for the galaxy power spectrum analysis indicates a broad agreement. We are grateful to the authors for coordinating with us the publication on the ArXiv website.

\section{Acknowledgements}

We thank Valentin Assassi, Daniel Baumann, Paolo Creminelli, Tommaso Giannantonio, Lam Hui, Ravi Sheth, Meng Su and Amit Yadav for helpful discussions and correspondence. We are grateful to the organizers of the ``String Theory and Precision Cosmology'' workshop held at Cornell University from the 25$^{th}$ to the 29$^{th}$ of July, 2011, where this project was initiated. CMB simulations were performed on the COSMOS supercomputer  which is funded by STFC, BIS and SGI. E.S. was supported in part by the EU Marie Curie Inter-European Fellowship. He is also grateful to the Department of Applied Mathematics and Theoretical Physics of the University of Cambridge for hospitality during the completion of this work. J.R.F. and E.P.S. were supported by STFC grant ST/F002998/1 and the Centre for Theoretical Cosmology. X.C. is supported by the Stephen Hawking Advanced Fellowship.

\end{document}